# Accurate and fast small *p*-value estimation for permutation tests in high-throughput genomic data analysis with the cross-entropy method

Yang Shi[1,5,7,*], Weiping Shi[2], Mengqiao Wang[3], Ji-Hyun Lee[4], Huining Kang[5,6,*] and Hui Jiang[7,8,9,*]

[1]Division of Biostatistics and Data Science, Department of Population Health Sciences and Department of Neuroscience and Regenerative Medicine, Medical College of Georgia, Augusta University, Augusta, Georgia 30912, United States

[2]College of Mathematics, Jilin University, Changchun, Jilin 130012, China

[3]Department of Epidemiology and Biostatistics, School of Public Health, Chengdu Medical College, Chengdu, Sichuan 610500, China

[4]Division of Quantitative Sciences, University of Florida Health Cancer Center and Department of Biostatistics, University of Florida, Gainesville, Florida 32610, United States

[5]University of New Mexico Comprehensive Cancer Center Biostatistics Shared Resource and [6]Department of Internal Medicine, University of New Mexico, Albuquerque, New Mexico 87131, United States

[7]Department of Biostatistics, [8]Center for Computational Medicine and Bioinformatics and [9]University of Michigan Rogel Cancer Center, University of Michigan, Ann Arbor, Michigan 48109, United States.

*Corresponding authors. Email addresses: yshi@augusta.edu (YS), hukang@salud.unm.edu (HK) and jianghui@umich.edu (HJ).

## Abstract

Permutation tests are widely used for statistical hypothesis testing when the sampling distribution of the test statistic under the null hypothesis is analytically intractable or unreliable due to finite sample sizes. One critical challenge in the application of permutation tests in genomic studies is that an enormous number of permutations are often needed to obtain reliable estimates of very small *p*-values, leading to intensive computational effort. To address this issue, we develop algorithms for the accurate and efficient estimation of small *p*-values in permutation tests for paired and independent two-group genomic data, and our approaches leverage a novel framework for parameterizing the permutation sample spaces of those two types of data respectively using the Bernoulli and conditional Bernoulli distributions, combined with the cross-entropy method. The performance of our proposed algorithms is demonstrated through the application to two simulated datasets and two real-world gene expression datasets generated by microarray and RNA-Seq technologies and comparisons to existing methods such as crude permutations and SAMC, and the results show that our approaches can achieve orders of magnitude of computational efficiency gains in estimating small *p*-values. Our approaches offer promising solutions for the improvement of computational efficiencies of existing permutation test procedures and the development of new testing methods using permutations in genomic data analysis.

**Keywords:** permutation test; the cross-entropy method; Monte Carlo simulation; importance sampling; *p*-value; genomic data analysis

## 1. Introduction

Permutation tests are a popular method for statistical hypothesis testing when the sampling distribution of the test statistic under the null hypothesis is analytically intractable or unreliable due to finite sample sizes. Compared to parametric testing methods that usually have distributional assumptions on the studied population and rely on the asymptotic distributions of the test statistics, permutation tests have less stringent assumptions, which only assume exchangeability of the observations under the null hypothesis and are relatively easy to implement in practice (Pesarin and Salmaso 2010). Therefore, they have wide applications in genomic studies nowadays (Tusher, Tibshirani, and Chu 2001; Segal et al. 2018; Browning 2008; Pahl and Schafer 2010; Che et al. 2014). However, a critical challenge for applying permutation tests is that the computational burden is intensive when small *p*-values are needed to be accurately estimated. This situation is very common in genomic studies where many tests are performed simultaneously and the family-wise error rate or the false-discovery rate needs to be controlled at an acceptable level. Consequently, the *p*-value of an individual test for a genomic feature of interest (e.g. a gene in an RNA-Seq experiment or a single nucleotide polymorphism, SNP, in a genome-wide association study, GWAS) needs to be small enough to achieve statistical significance. For instance, in a differential gene expression (DGE) analysis with over 20,000 genes, usually a *p*-value less than $10^{-5}$ to $10^{-6}$ needs to be achieved for a gene to be declared as significantly differential expressed (Tusher, Tibshirani, and Chu 2001); in a GWAS with half a million SNPs, usually a *p*-value less than $10^{-7}$ needs to be achieved for a SNP to be declared as genome-widely significant (Yu et al. 2011). In addition, in genomic studies such as DGE analysis and GWAS, it is desirable to rank the statistically significant signals by their *p*-values (often together with their effect sizes) so that the researchers can prioritize and follow up with those significant genomic features for further biological studies, which also requires that the small *p*-values associated with those top significant signals to be reliably estimated. It is not rare to see very small *p*-values are reported in published DGE and GWAS studies (Bangalore, Wang, and Allison 2009).

The goal of this work is to improve the efficiency of computing small *p*-values in permutation tests. To introduce our methodology, it is helpful to situate permutation tests within the broader context of Monte Carlo (MC) sampling methods Monte Carlo (MC) sampling methods, which first generate a large number of resamples via random permutations (i.e. sampling without replacement) of the observed data and then repeatedly calculate the test statistics using those resamples and estimate the *p*-value as the proportion of the test statistics based on the MC samples that are more extreme than the one based on the observed data. Under such framework, estimating a small *p*-value is essentially estimating the probability of a rare event in MC simulations. The cross-entropy (CE) method is an efficient and powerful algorithm for estimating the small probability of a rare event in MC simulations and has been widely used in the field of operations research (Rubinstein and Kroese 2004; Rubinstein 1997). It relies on the importance sampling (IS) technique and its basic idea is to find a proposal density, called the CE-optimal proposal density, that approximates the optimal proposal distribution in IS adaptively by minimizing the CE between them. Regarding the application of the CE method in biostatistics, Hu and Su first develop an algorithm for efficient estimation of the tail probabilities and quantiles of non-parametric bootstrapping procedures (note that the bootstrap is another MC-based method for statistical inference that generates the resamples via sampling the observed data with replacement (Efron and Tibshirani 1994; Davison and Hinkley 1997)), and they show that more than an order of magnitude of computational efficiency gains can be achieved via the CE method (Hu and Su 2008a, 2008b). More recently, some progress has been made to extend the CE method. The so-called improved CE algorithm is proposed in (Chan and Kroese 2012), which directly draws random samples from the optimal proposal distribution of IS with Markov chain Monte Carlo techniques such as the Gibbs sampler and then estimates the CE-optimal proposal distribution using those samples. In that way, the CE-optimal proposal distribution can be obtained in a single step by avoiding adaptively and iteratively updating the parameters, thus it achieves higher estimation precision and computational efficiency than the adaptive CE method. Based on the principle of improved CE algorithm (Chan and Kroese 2012), a new algorithm was developed in (Shi et al. 2019) that uses the Hamiltonian Monte Sampler to draw random samples from the optimal proposal distribution of IS and applied this algorithm to efficiently estimate small *p*-values for a few widely used statistics in genomics such as the quadratic or constrained linear functions of multivariate normal random variables.

However, those existing methods including (Shi et al. 2019; Chan and Kroese 2012) cannot be directly applied to estimate the small *p*-values in permutation tests, because they require that the data of interest comes from a fully parameterized probability distribution and the probability density or mass of that distribution can be evaluated pointwisely. To the best of our knowledge, there is no existing approach for the parameterization of the sample space of permutation tests. The main contribution of this work is that we devise novel approaches for parameterizing the permutation sample space, which make it feasible for the application of the CE method for estimating small *p*-values in permutation tests. Specifically, we aim to estimate the small *p*-values in permutation tests for paired and independent two-group data, and we propose that the permutation sample space of such two types of data can be respectively parameterized by the joint i.i.d. Bernoulli distributions and the conditional Bernoulli distribution. Under such parameterizations, estimating small *p*-values from permutation tests can be embedded into the framework of importance sampling where the optimal proposal distribution can be obtained by the CE method. By combining the parameterization of the permutation sample space and the CE method, we can achieve orders of magnitude of computational efficiency gains in estimating small *p*-values in permutation tests compared to crude permutations. The rest parts of this paper are organized as follows: First, we briefly introduce the CE method, and then discuss in detail how the permutation sample spaces of paired and independent two-group permutation tests can be parameterized and present our algorithms for estimating small *p*-values in permutation tests. Next, we demonstrate the performance of our proposed approach by applying it to two simulated datasets and two real-world gene expression datasets generated by microarray and RNA-Seq experiments and comparing it to existing methods. Finally, discussions on recommendations for implementing those algorithms in practice and possible extensions based on our current proposed framework are given.

## 2. Methods

### 2.1 Monte Carlo and the cross-entropy method

Since our approach is based on the adaptive CE method for estimating small probabilities, here we briefly introduce the adaptive CE method following (Shi et al. 2019; Rubinstein and Kroese 2004). Suppose that we are performing a hypothesis test that testing the null hypothesis $H_0$ versus the alternative hypothesis $H_1$, and further assume that $H_0$ is a simple null hypothesis for simplicity, then the *p*-value from the test is

$$p = Pr[T(\mathbf{Z}) \geq \gamma \mid H_0], \tag{1}$$

where $\mathbf{Z}$ is the sample, $T(\cdot)$ is the test statistic used in the test that is some real function of $\mathbf{Z}$ measuring the discrepancy between the data and the null hypothesis, and $\gamma$ is the test statistic calculated from the observed data (Davison and Hinkley 1997). The symbol of conditioning on $H_0$ will be dropped for simplicity hereafter. If we can assume that the sample $\mathbf{Z}$ comes a fully parameterized probability distribution $f(\cdot;\boldsymbol{\theta}_0)$ with $\boldsymbol{\theta}_0$ as the parameter vector, and random samples can be generated from $f(\cdot;\boldsymbol{\theta}_0)$, the *p*-value defined in (1) can be estimated by MC simulations as

$$\hat{p} = \frac{1}{N}\sum_{l=1}^{N} I\{T(\mathbf{z}_l) \geq \gamma\}, \tag{2}$$

where $\mathbf{z}_1,...,\mathbf{z}_N$ are random samples drawn from $f(\cdot;\boldsymbol{\theta}_0)$, $I(\cdot)$ is the indicator function and the form of $\hat{p}$ as in (2) is called the MC estimator of *p*. The problem to be addressed is when *p* is very small, accurately estimating *p* using the crude MC estimator in (2) is very computationally intensive or even infeasible when $p$ is extremely small. The adaptive CE method is a generic algorithm for efficiently estimating such small probabilities in MC simulations. Let $\mathcal{Z}$ be the sample space of $\mathbf{Z}$, then the $p$-value defined in (1) can be further written as

$$p = Pr[T(\mathbf{Z}) \geq \gamma] = E_{\boldsymbol{\theta}_0}[I\{T(\mathbf{Z}) \geq \gamma\}] = \int_{\mathcal{Z}} I\{T(\mathbf{Z}) \geq \gamma\} f(\mathbf{Z};\boldsymbol{\theta}_0)\mu(d\mathbf{Z}), \quad (3)$$

where the expectation $E$ is taken with respect to $f(\cdot;\boldsymbol{\theta}_0)$ and $\mu$ is the probability measure on which $f(\cdot;\boldsymbol{\theta}_0)$ is based. Below we take $\mu(d\mathbf{Z}) = d\mathbf{Z}$ and drop $\mathcal{Z}$ for simplicity. The CE method relies on the IS approach. Let $g(\cdot)$ be the proposal density used in IS, then (3) can be written as

$$p = \int I\{T(\mathbf{Z}) \geq \gamma\} \frac{f(\mathbf{Z};\boldsymbol{\theta}_0)}{g(\mathbf{Z})} g(\mathbf{Z})\, \mathrm{d}\mathbf{Z} = E_g[I\{T(\mathbf{Z}) \geq \gamma\} \frac{f(\mathbf{Z};\boldsymbol{\theta}_0)}{g(\mathbf{Z})}], \quad (4)$$

where the subscript $g$ denotes that the expectation is taken with respect to $g(\cdot)$ now. Then $p$ can be estimated by the MC estimator of (4) as

$$\hat{p} = \frac{1}{N} \sum_{l=1}^{N} I\{T(\mathbf{z}_l) \geq \gamma\} \frac{f(\mathbf{z}_l;\boldsymbol{\theta}_0)}{g(\mathbf{z}_l)}, \quad (5)$$

where $\mathbf{z}_l$ 's, $l = 1, \ldots, N$, are random samples drawn from $g(\cdot)$. There is an optimal proposal density under which the IS estimator (5) has zero variance, which is

$$g^*(\mathbf{Z}) = \frac{I\{T(\mathbf{Z}) \geq \gamma\} f(\mathbf{Z};\boldsymbol{\theta}_0)}{p}. \quad (6)$$

However, $g^*$ cannot be directly used as the proposal density for estimating $p$ in (5), since it contains the unknown probability $p$ that is the quantity to be estimated. The CE method provides a general solution to finding a proposal distribution $f(\cdot;\boldsymbol{\theta})$ which approximates the optimal proposal distribution $g^*$ within the same distribution family as $f(\cdot;\boldsymbol{\theta}_0)$ in the sense that the Kullback–Leibler divergence (a.k.a. the Kullback–Leibler cross-entropy) between $f(\cdot;\boldsymbol{\theta})$ and $g^*$ is minimized:

$$\begin{aligned} D\left(g^*(\cdot), f\left(\cdot;\boldsymbol{\theta}\right)\right) &:= \int g^*(\mathbf{Z}) \ln \frac{g^*(\mathbf{Z})}{f\left(\mathbf{Z};\boldsymbol{\theta}\right)} \mathrm{d}\mathbf{Z} \\ &= \int g^*(\mathbf{Z}) \ln g^*(\mathbf{Z})\, \mathrm{d}\mathbf{Z} - \int g^*(\mathbf{Z}) \ln f\left(\mathbf{Z};\boldsymbol{\theta}\right) \mathrm{d}\mathbf{Z}, \end{aligned} \quad (7)$$

where $f(\cdot;\boldsymbol{\theta})$ is referred as the CE-optimal proposal distribution, and it is the proposal density that the CE method tries to find. Since the first term on the right-hand side of the second equality in (7) does not contain $\boldsymbol{\theta}$, hence the parameter $\boldsymbol{\theta}$ that minimizes $D\left(g^*(\cdot), f\left(\cdot;\boldsymbol{\theta}\right)\right)$ should maximize the second term, which can be further expressed as

$$\begin{aligned}\int g^{*}(\mathbf{Z})\ln f\left(\mathbf{Z};\boldsymbol{\theta}\right)\mathrm{d}\,\mathbf{Z} &= \int \frac{I\{T(\mathbf{Z})\geq\gamma\}f(\mathbf{Z};\boldsymbol{\theta}_0)}{p}\ln f\left(\mathbf{Z};\boldsymbol{\theta}\right)\mathrm{d}\,\mathbf{Z} \\ &= \frac{1}{p}E_{\boldsymbol{\theta}_0}[I\{T(\mathbf{Z})\geq\gamma\}\ln f\left(\mathbf{Z};\boldsymbol{\theta}\right)].\end{aligned} \tag{8}$$

Therefore, $\boldsymbol{\theta}$ should maximize the expectation in (8), which can be rewritten as

$$\arg\max_{\boldsymbol{\theta}} E_{\boldsymbol{\theta}_0}[I\{T(\mathbf{Z})\geq\gamma\}\ln f\left(\mathbf{Z};\boldsymbol{\theta}\right)] = \arg\max_{\boldsymbol{\theta}} E_{\boldsymbol{\theta}_t}[I\{T(\mathbf{Z})\geq\gamma\}\frac{f(\mathbf{Z};\boldsymbol{\theta}_0)}{f(\mathbf{Z};\boldsymbol{\theta}_t)}\ln f(\mathbf{Z};\boldsymbol{\theta})], \tag{9}$$

where the equality resulted from using IS again with a change of the proposal density to $f(\cdot;\boldsymbol{\theta}_t)$, another distribution within the same family as $f(\cdot;\boldsymbol{\theta}_0)$. The solution to the right-hand side of the equality in (9) can be found by solving the following

$$\arg\max_{\boldsymbol{\theta}} \frac{1}{N}\sum_{l=1}^{N}[I\{T(\mathbf{z}_l)\geq\gamma\}\frac{f(\mathbf{z}_l;\boldsymbol{\theta}_0)}{f(\mathbf{z}_l;\boldsymbol{\theta}_t)}\ln f(\mathbf{z}_l;\boldsymbol{\theta})], \tag{10}$$

where $\mathbf{z}_l$'s, $l$ = 1, ..., $N$, are random samples drawn from $f(\cdot;\boldsymbol{\theta}_t)$ and (10) is called the stochastic counterpart of the right-hand side of the equality in (9). When the probability of the target event $I\{T(\mathbf{z}_l)\geq\gamma\}$ is small, (10) can be solved iteratively using the following adaptive CE method.

**Algorithm 0:** The adaptive CE method for rare-event probability estimation

**A. Adaptive parameter-updating step:**

(1) Specify a constant $\rho\in\left(0,1\right)$. Start with $\boldsymbol{\theta}_0$; Set the iteration counter $t$ = 0.

(2) At the $t$th iteration, generate $N$ random samples $\mathbf{z}_1,\ldots,\mathbf{z}_N$ from $f(\cdot;\boldsymbol{\theta}_t)$. Calculate the statistics $T(\mathbf{z}_1),\ldots,T(\mathbf{z}_N)$, and compute $\gamma_t$ as their $(1-\rho)$ sample quantile provided $\gamma_t\leq\gamma$. If $\gamma_t>\gamma$, set $\gamma_t=\gamma$.

(3) Update the parameter $\boldsymbol{\theta}_t$ with $\boldsymbol{\theta}_{t+1}$, which is the solution to the following problem:

$$\boldsymbol{\theta}_{t+1} = \arg\max_{\boldsymbol{\theta}} \frac{1}{N}\sum_{l=1}^{N}[I\{T(\mathbf{z}_l)\geq\gamma_t\}\frac{f(\mathbf{z}_l;\boldsymbol{\theta}_0)}{f(\mathbf{z}_l;\boldsymbol{\theta}_t)}\ln f(\mathbf{z}_l;\boldsymbol{\theta})] \tag{11}$$

where $\mathbf{z}_l$'s, $l$ = 1, ..., $N$, are random samples drawn from the proposal density $f(\cdot;\boldsymbol{\theta}_t)$, and (11) will be referred as **the CE formula** throughout the rest of the paper.

(4) If $\gamma_t\leq\gamma$, set $t$ = $t$ + 1 and reiterate from above Step (2); else, proceed to the following estimating step.

**B. Estimating step:**

Use $f(\cdot;\boldsymbol{\theta}_t)$ as the proposal distribution and generate random samples $\mathbf{z}_1,...,\mathbf{z}_M$ from $f(\cdot;\boldsymbol{\theta}_t)$. Estimate $p$ as $\hat{p}=\frac{1}{M}\sum_{l=1}^{M}[I\{T(\mathbf{z}_l)\geq\gamma\}\frac{f(\mathbf{z}_l;\boldsymbol{\theta}_0)}{f(\mathbf{z}_l;\boldsymbol{\theta}_t)}]$.

The rationale of the adaptive CE method is that a sequence of updated parameter values $\{\boldsymbol{\theta}_t, t=0,1...\}$ and threshold values $\{\gamma_t, t=0,1...\}$ are generated in Step A, and it can be shown that $\{\gamma_t, t=0,1...\}$ is monotonically non-decreasing under mild regularity conditions and the stopping condition $\gamma_t>\gamma$ can be reached with high probability in a finite number of iterations for small $\rho$ (Rubinstein and Kroese 2004; Rubinstein 1999). Hence, the updated parameters $\{\boldsymbol{\theta}_t, t=0,1...\}$ will be getting closer and closer to the optimal parameter $\boldsymbol{\theta}$ that we want to find in problem (10). The estimating step is just a standard IS step that uses $f(\cdot;\boldsymbol{\theta}_t)$ as the proposal distribution.

**2.2 Parameterization of permutation sample space and application of the CE method**

The adaptive CE method has two requirements: (1) The sample $\mathbf{Z}$ is from a fully parameterized probability distribution $f(\cdot;\boldsymbol{\theta}_0)$ and $f(\cdot;\boldsymbol{\theta}_0)$ can be evaluated pointwisely for each sample $\mathbf{z}_l$ from this distribution; (2) Random samples can be generated from the family of distributions that $f(\cdot;\boldsymbol{\theta}_0)$ belongs to. Below we show how the permutation sample spaces for paired and independent two-group permutation tests can be parameterized to satisfy those requirements.

2.2.1 Paired two-group data

Testing a location shift of paired two-group data is equivalently testing whether the differences between the paired observations are symmetrically distributed around 0 (Pesarin and Salmaso 2010). Let $\mathbf{x}=[x_1,...,x_n]^T$ be the observed data of the differences between the paired observations. A widely-used crude permutation procedure generates the permutated samples by randomly assigning the + or − sign to $x_i$'s, $i=1,...,n$, with equal probability of 1/2 (Pesarin and Salmaso 2010). Let $\mathbf{z}_l=[z_{l1},z_{l2},...,z_{ln}]^T$, $l=1,...,N$, be the permuted samples generated by this procedure, where $N$ is the number of permutations. To parameterize the permutation sample space, define a binary variable $s_{li}$ taking values on 0 or 1 that indicates the sign assigned to $x_i$ for the $l$th permuted sample with 1 meaning + and 0 meaning −. Further, define $p_i$'s, $i=1,...,n$, as the probability of assigning the + sign to $x_i$. Therefore, $s_{li}$ follows a Bernoulli distribution with probability $p_i$, where

$$s_{li} = \begin{cases} 1, \text{ with probability } p_i \\ 0, \text{ with probability } 1 - p_i \end{cases}. \tag{12}$$

Let $\mathbf{s}_l = [s_{l1},...,s_{ln}]^T$ and $\mathbf{p} = [p_1,..., p_n]^T$ be the vector forms of $s_{li}$'s and $p_i$'s, respectively (we will drop the subscript $l$ below if there is no ambiguity). Hence, the permutation sample space of paired two-group data can be parameterized by the joint distribution of $\mathbf{s}_l$, which is the joint distribution of $n$ i.i.d. Bernoulli variables with probability vector $\mathbf{p}$ and the density

$$f(\mathbf{z};\mathbf{p}) = f(\mathbf{s};\mathbf{p}) = \prod_{i=1}^{n}[p_i^{s_i}(1-p_i)^{1-s_i}]. \tag{13}$$

Using (13), we can update the probability vector $\mathbf{p}$ via the adaptive CE method (Algorithm 0) with starting value $\mathbf{p}_0 = [1/2,...,1/2]^T$ under the crude permutation procedure. Substituting $f(\cdot;\cdot)$ in (11) with (13) gives the following optimization problem

$$\mathbf{p}_{t+1} = \underset{\mathbf{p}=[p_1,...,p_n]^T}{\arg\max} \frac{1}{N}\sum_{l=1}^{N}[I\{T(\mathbf{z}_l) \geq \gamma_t\}Q(\mathbf{z}_l;\mathbf{p}_0,\mathbf{p}_t)\ln f(\mathbf{z}_l;\mathbf{p})], \tag{14}$$

where $Q(\mathbf{z}_l;\mathbf{p}_0,\mathbf{p}_t) = \dfrac{f(\mathbf{z};\mathbf{p}_0)}{f(\mathbf{z};\mathbf{p}_t)}$ is the likelihood ratio, and $\mathbf{p}_t$ and $\mathbf{p}_{t+1}$ are the updated parameter vectors at the $t$th and ($t$+1)th iteration, respectively. Problem (14) can be solved analytically by the following formula with details derived in Appendix A1:

$$p_{i,t+1} = \frac{\sum_{l=1}^{N}[I\{T(\mathbf{z}_l) \geq \gamma_t\}Q(\mathbf{z}_l;\mathbf{p}_0,\mathbf{p}_t)s_{li}]}{\sum_{l=1}^{N}[I\{T(\mathbf{z}_l) \geq \gamma_t\}Q(\mathbf{z}_l;\mathbf{p}_0,\mathbf{p}_t)]}, \tag{15}$$

where $p_{i,t+1}$, $i = 1, \ldots, n$, is the $i$th element of $\mathbf{p}_{t+1}$. Combining (13) to (15) and Algorithm 0, we have the following algorithm for paired two-group data.

**Algorithm 1:** The CE algorithm for estimating small $p$-values from paired two-group permutation test **(CE-Perm1)**

**A. Adaptive parameter-updating step:**

(1) Specify a small constant $\rho \in (0,1)$. Start with the initial probability vector $\mathbf{p}_0 = [1/2,...,1/2]^T$. Set the iteration counter $t = 0$.

(2) At the *t*th iteration, generate *N* random samples $\mathbf{z}_1,...,\mathbf{z}_N$ from *n* i.i.d. Bernoulli distributions with density (13). Calculate the statistics $T(\mathbf{z}_1),\ldots,T(\mathbf{z}_N)$, and compute $\gamma_t$ as their $(1-\rho)$ sample quantile, provided $\gamma_t \le \gamma$. If $\gamma_t > \gamma$, set $\gamma_t = \gamma$.

(3) Update the parameter $\mathbf{p}_t$ as $\mathbf{p}_{t+1}$ using (15).

(4) If $\gamma_t \le \gamma$, set $t = t + 1$ and reiterate from above step (2); otherwise, proceed to the following estimating step.

**B. Estimating step:**

Use $f(\cdot;\mathbf{p}_t)$ as the proposal distribution and generate *M* random samples $\mathbf{z}_1,...,\mathbf{z}_M$ from $f(\cdot;\mathbf{p}_t)$.

Estimate the *p*-value as $\hat{p} = \frac{1}{M}\sum_{l=1}^{M}[I\{T(\mathbf{z}_l) \ge \gamma\}\frac{f(\mathbf{z}_l;\mathbf{p}_0)}{f(\mathbf{z}_l;\mathbf{p}_t)}]$.

2.2.2 Permutation test for independent two-group data

Following similar notations above, let $\mathbf{x} = [x_1,...,x_n]^T$ be the observed data of the two groups and $\mathbf{z}_l = [z_{l1},...,z_{ln}]^T$, $l = 1,...,N$, be the permuted samples, where *N* is the number of permutations. To assign group labels to the data, without loss of generality, we assume that the first *k* elements of $\mathbf{x}$ belong to Group 1 and the remaining $m = n - k$ elements of $\mathbf{x}$ belong to Group 2 with $0 < k \le m$.

To parameterize the permutation sample space for independent two-group data, we propose to use the conditional Bernoulli (CB) distribution (Chen 2000; Chen and Liu 1997; Chen, Dempster, and Liu 1994). First, let $\mathbf{d}_l = [d_{l1},...,d_{ln}]^T$ be a **partition vector**, where $d_{li}$'s, $i = 1,...,n$, takes a value of either 1 or 0 with 1 indicating that $x_i$ belongs to Group 1 and 0 indicating Group 2 in the permuted sample $\mathbf{z}_l$ (e.g. suppose $n = 6$, $k = 2$ and $m = 4$, then $\mathbf{d}_l = [1,0,1,0,0,0]^T$ means that $\{x_1, x_3\}$ belong to Group 1 and $\{x_2, x_4, x_5, x_6\}$ belong to Group 2 in the *l*th permuted sample). We will drop the subscript *l* hereafter if there is no ambiguity. The conditional distribution of $\mathbf{d} = [d_1, d_2,...,d_n]^T$ with $d_i \sim Bernoulli(p_i)$ and $\sum_{i=1}^{n} d_i = k,\ k = 1,...,n$, is called the CB distribution with density

$$f(\mathbf{d};\mathbf{w}) = \Pr(d_1, d_2,...,d_n \mid \sum_{i=1}^{n} d_i = k) = \frac{\prod_{i=1}^{n} w_i^{d_i}}{R_k}, \tag{16}$$

where

$$R_k = \Pr(\sum_{i=1}^{n} d_i = k)\prod_{i=1}^{n}(1+w_i)$$

is a normalization constant and $\mathbf{w} = [w_1,...,w_n]^T$, $w_i = p_i/(1-p_i)$, $i = 1,...,n$, is the vector of odds (Kroese, Taimre, and Botev 2011; Chen 2000). Note that $w_i$'s (or equivalently, $p_i$'s) are the parameters of the CB distribution, and we have $\mathbf{w} = [1,...,1]^T$ (or equivalently all $p_i$'s equal to 1/2) under the crude permutation procedure where the permutated samples are randomly generated. Our aim is to update the parameter vector $\mathbf{w}$ using the adaptive CE method (Algorithm 0). To achieve that, we need to address the following two questions:

(1). How to efficiently generate random samples from the CB distribution? To that end, five algorithms are developed in the literature (Chen and Liu 1997; Chen, Dempster, and Liu 1994). Here we use the **drafting sampling algorithm** (Chen and Liu 1997; Chen, Dempster, and Liu 1994), which is given as Procedure A2 with details in Appendix A2.

(2). How to efficiently optimize the CE formula (11) parameterized by the CB distribution (15)? To solve this problem, an iterative algorithm is given as Procedure A3 with details in Appendix A3.

Combining the Procedure A2, A3 and Algorithm 0, we have the following algorithm for independent two-group data.

**Algorithm 2:** The CE algorithm for estimating small *p*-values from independent two-group permutation test **(CE-Perm2)**

**A. Adaptive parameter-updating step:**

(1) Specify a constant $\rho \in (0,1)$. Start with the initial parameter vector $\mathbf{w}_0 = [1,...,1]^T$ (equivalently, $\mathbf{p}_0 = [1/2,...,1/2]^T$). Set the iteration counter $t = 0$.

(2) At the $t$th iteration, generate $N$ random samples $\mathbf{z}_1,...,\mathbf{z}_N$ from the CB distribution $f(\cdot;\mathbf{w}_t)$ using the drafting sampling algorithm (Procedure A2 in Appendix A2). Calculate the statistics $T(\mathbf{z}_1),\ldots,T(\mathbf{z}_N)$, and compute $\gamma_t$ as their $(1-\rho)$ sample quantile, provided $\gamma_t \le \gamma$. If $\gamma_t > \gamma$, set $\gamma_t = \gamma$.

(3) Update the parameter $\mathbf{w}_t$ as $\mathbf{w}_{t+1}$ using Procedure A3 in Appendix A4.

(4) If $\gamma_t \le \gamma$, set $t = t + 1$ and reiterate from above step (2); otherwise, proceed to the following estimating step.

**B. Estimating step:**

Use $f(\cdot;\mathbf{w}_t)$ as the proposal distribution and generate $M$ random samples $\mathbf{z}_1,...,\mathbf{z}_M$ from $f(\cdot;\mathbf{w}_t)$.

Estimate the $p$-value as $\hat{p}=\frac{1}{M}\sum_{l=1}^{M}\left[I\{T(\mathbf{z}_l)\geq\gamma\}\frac{f(\mathbf{z}_l;\mathbf{w}_0)}{f(\mathbf{z}_l;\mathbf{w}_t)}\right]$.

### 2.3 Variance reduction with the screening method

One limitation of the CE-Perm1 and CE-Perm2 algorithms is that the variance of the estimated $p$-values grow with the dimensionality of the parameter vectors (i.e. $\mathbf{p}$ in CE-Perm1 and $\mathbf{w}$ in CE-Perm2), and this issue is further elaborated in the Summary and Discussion section. To improve this problem, a modified version of the screening method proposed in (Rubinstein and Glynn 2009) is used in both algorithms to further reduce the variance in high-dimensional parameter spaces, and the two resulting improved algorithms are given as Algorithm 1S (CE-Perm1-S) and Algorithm 2S (CE-Perm2-S) in Appendix A4.

## 3. Results

### 3.1 Simulation studies

Simulations are used to assess the accuracy, variations and efficiency of the proposed algorithms. Ideally, we hope to compare the variances of the estimates and the computational time from our algorithms and those from crude permutations using the same test statistics for a range of small $p$-values with different orders of magnitude. However, for those very small $p$-values (e.g. less than $10^{-9}$), crude permutations will take unaffordable time and memories to obtain an accurate estimate. Therefore, in the paired and independent two-group data studied below, we use the differences of the two sample means as the test statistics for the permutation tests and generate the data from normal distributions, where the student's $t$-test is equivalent to the permutation tests asymptotically for comparing the means (Lehmann and Romano 2005; Segal et al. 2018) and thus should give similar $p$-values as the permutation tests. In this way, we can use the $p$-values from the $t$-test to estimate the orders of magnitude and check the reliance of those from the permutation tests, though it should be noted that the two different types of tests will not give exactly the same $p$-values due to the finite sample sizes used in the simulations.

#### 3.1.1 Paired two-group example

Since comparing the means of paired two-group data is equivalently testing whether the mean of the differences between the paired observations equal to 0, we generate a sequence of one-group random samples with sample size $n = 100$ from $N(\mu, 1)$ [ $N(\mu, \sigma)$ denotes a normal distribution with mean $\mu$ and standard deviation $\sigma$ ] with the effect size $\mu$ varying from 0.275 to 0.725, and then apply the crude permutation procedure, CE-Perm1 and CE-Perm1-S to test whether the means equal to 0. The test statistic *T* used here is the sample mean. For crude permutation test, the number of permutated samples used, $N_p$, is determined as $N_p = 100 / \tilde{p}$, where $\tilde{p}$ is the order of magnitude of the *p*-value estimated by the one-sample *t*-test (Table 1). The crude permutations are only run for *p*-values with orders of magnitude from $10^{-5}$ to $10^{-8}$ as the computation time is unaffordable for those less than $10^{-8}$. For CE-Perm1, the parameter $\rho$ is set as 0.1, the number of random samples used in each iteration of the adaptive parameter-updating step (*N*) is 2000 and the number of random samples used in the estimating step (*M*) is 10000. For CE-Perm1-S, the number of repetitions in the parameter-screening step is set as 10 and the parameter setting of the rest steps is the same as CE-Perm1. To obtain reliable estimation of the *p*-value and its variations, each method is repeated 100 times for each single *p*-value. For each method, the sample mean of the estimated *p*-values from the 100 runs is used as the final point estimate of the *p*-value and the Monte Carlo relative error (*MCRE*) (Kroese, Taimre, and Botev 2011), defined as $MCRE = S / \hat{p}\sqrt{N_r}$, is used to assess the variations, where $N_r$ is the number of repeated runs of each method (i.e. 100 here), $\hat{p}$ and *S* are the mean and standard deviation of the estimated *p*-values from the 100 runs.

Table 1 shows the estimated results and the computation time of each method. For those *p*-values for which the crude permutations are affordable, the point estimates of the *p*-values and their variations from all three methods are similar, while CE-Perm1 and CE-Perm1-S reduce the computation time by roughly a factor from 100 to 79000 compared to the crude permutations and the gain of efficiency increases as the *p*-value goes smaller (Table 1, *p*-values from $10^{-5}$ to $10^{-8}$). CE-Perm1 and CE-Perm1-S are also fast for *p*-values from $10^{-9}$ to $10^{-15}$ with CPU times less than 2 minutes. CE-Perm1-S roughly reduces the *MCRE*s between 20% and 50% compared to CE-Perm1 (Table 1).

3.1.2 Independent two-group example

We provide two examples with different sample sizes. In the first example, we test the means of two groups with equal sample sizes of 20. The data of the first group are generated from $N(0,1)$ and the data of the second group are generated from $N(\mu,1)$ with effect size $\mu$ varying from 1 to 3.25, and then a sequence of *p*-values with orders of magnitude from $10^{-5}$ to $10^{-15}$ are obtained by comparing the means of the two groups (similar to Section 3.1.2, the orders of magnitude of them can be estimated by the two-sample *t*-test, see Table 2). Then crude permutations, CE-Perm2 and CE-Perm2-S are used and compared. Also included in the comparisons is another algorithm called SAMC implemented in the R package *EXPERT*, which has a similar goal to our method but uses the stochastic approximation Markov chain Monte Carlo (MCMC) algorithm (Yu et al. 2011) and can be only applied to independent two-group comparison with its current implementation. The test statistic *T* used is the difference of the sample means between the two groups. The number of permutations used in the crude permutations and the parameter values used in CE-Perm2 and CE-Perm2-S are the same as those in Section 3.1.1. For SAMC, we use default values of the program, which is $2\times10^5$ resamples in the initial step for refining the partitions of the test statistic and $10^6$ resamples in the final step for estimating each *p*-value. Same as in Section 3.1.1, each method is repeated 100 times, and the average of the estimated *p*-values from the 100 runs is used as the final point estimate of the *p*-value and the *MCRE* is used to assess the variations.

Table 2 shows the results and the computation time of each method in this example. The estimated *p*-values from all methods and their variations are similar (crude permutation test is only run for *p*-values from $10^{-5}$ to $10^{-8}$), where CE-Perm2 and CE-Perm2-S give smaller *MCRE*s than the crude permutations and SAMC. In terms of computation time, CE-Perm2 and CE-Perm2-S reduce it by roughly a factor from 25 to 8700 compared to the crude permutations and the gain of efficiency increases as the *p*-value goes smaller (Table 2, *p*-values from $10^{-5}$ to $10^{-8}$). We note that the SAMC algorithm is implemented by incorporating C codes in R in the *EXPERT* package while CE-Perm2 and CE-Perm2-S are purely implemented in R, so the computation time between them may not be directly comparable. Despite of that, CE-Perm2 and CE-Perm2-S show substantial gain in computational efficiency compared to SAMC in this example (Table 2).

We also present another example with the sample sizes of the two groups equal to 100 and the orders of magnitude of *p*-values to be estimated ranges from $10^{-5}$ to $10^{-15}$. Here we run SAMC with different number of resamples: SAMC-I - we use $2\times10^5$ resamples in the initial step for refining the partitions of the test statistic and $5\times10^6$ resamples in the final step for estimating the *p*-value; SAMC-II - we use $2\times10^5$ resamples in the initial step for refining the partitions of the test statistic and $10^6$ permutated samples in the final step for estimating the *p*-value. For CE-Perm2, the parameter $\rho$ is set as 0.1 and $N = 4000$ random samples is used in each iteration of the adaptive parameter-updating step and $M = 20000$ random samples is used in the estimating step; for CE-Perm2-S, the number of repetitions in the parameter-screening step is set as 10, and the parameter setting of the rest steps is the same as CE-Perm2. Similarly, each method is repeated 100 times. The results of this example are shown in Table 3. The estimated *p*-values from all methods are similar. In terms of variations, though the *MCRE*s from the four methods are on the same orders of magnitude, the performance of CE-Perm2 is worse than the previous example with smaller sample size. This issue is known as the degeneracy of the likelihood ratios for IS in high dimensions (Rubinstein and Glynn 2009), which is the limitation of the adaptive CE method and will be further discussed in Section 4. CE-Perm2-S substantially alleviates this problem (Table 3). Regarding the computation time, CE-Perm2 and CE-Perm2-S are faster than SAMC I and II (Table 3).

### 3.2 Applications

We present two examples of the application of our algorithms to differential gene expression analyses of an RNA-Seq dataset (paired two-group) and a microarray dataset (independent two-group), respectively. The test statistics $T$ used are the moderated t-statistics for paired and independent two group data proposed in the *samr* package with their formulas given below (Tusher, Tibshirani, and Chu 2001). The sampling distributions of the moderated t-statistics under the null hypothesis that there is no differential expression of the gene (i.e. the two means are equal) are analytical intractable, and *samr* uses a permutation procedure to estimate the *p*-values and FDRs. One limitation of *samr*'s permutation procedure is that the *p*-values and FDRs are often estimated to be 0's for those top significant genes as presented in the two examples below, therefore we apply our algorithms to accurately estimate the *p*-values associated with those top genes and rank them by the estimated *p*-values.

#### 3.2.1 Application to an RNA-Seq dataset - paired two-group data

This RNA-Seq dataset is from the Cancer Genome Atlas (TCGA) project, where the expression levels of 20531 genes in the tumor tissue and the adjacent normal tissue to the tumor are measured by RNA-Seq for 51 patients with lung squamous cell carcinoma (LUSC). The goal is to identify which genes are significantly changed between the tumor and its adjacent normal tissue, therefore the design of this study is a paired two-group comparison. The test statistic $T$ used is the moderated $t$-statistic for paired two-groups proposed in *samr*,

$$T_{paired} = \frac{\sum_{i=1}^{n} d_{gi} / n}{s_g + s_0},$$

where $g = 1,...,G$ denotes the $g$th gene, $i = 1,...,n$ denotes the $i$th subject, $d_{gi}$'s are the differences in the expression levels of gene $g$ between the tumor and its adjacent normal tissue in the $i$th subject, $s_g$ is the standard error of $\sum_{i=1}^{n} d_{gi} / n$ as computed in the standard one-sample $t$-test, and $s_0$ is an exchangeability factor computed as a percentile of $s_g$'s, $g = 1,...,G$, that is, the standard errors of all genes (see the manual of *samr* for details). We first run the permutation procedure in *samr* with the number permutations set as 10000, and based on the results from *samr*, 41 genes are identified as the top significantly differentially expressed ones as their $p$-values and FDRs are smallest among all genes tested and are estimated to be 0's. To provide more accurate estimates of the $p$-values for those genes and rank them accordingly, we apply CE-Perm1 and CE-Perm1-S. For comparisons, we also perform crude permutations to estimate the p-values for those genes with an increased number of permutations of $10^8$. For CE-Perm1, the parameters are set as $\rho = 1$ and $N = 4000$ in the adaptive parameter-updating step and $M = 20000$ in the estimating step; for CE-Perm1-S, the number of repetitions in the parameter-screening step is set as 10, and the parameter settings in the adaptive parameter-updating and the estimating steps are the same as CE-Perm1. Each procedure is repeated 100 times for each individual gene. The sample mean of the estimated $p$-values from the 100 runs is used as the final point estimate of the $p$-value and the *MCRE* and standard deviation (SD) from the 100 runs are used to assess the variations. Supplementary Table S1 presents the results from all methods for the 41 genes and Table 4 presents the results from CE-Perm1 and CE-Perm1-S for the top 10 most significant genes. The results show that CE-Perm1 and CE-Perm1-S can accurately and efficiently estimate the $p$-values to the order of magnitude of $10^{-15}$ (Table 4 and Supplementary Table S1). Regarding the computation time, CE-Perm1 and CE-Perm1-S are about 650 times faster than the crude permutation procedure (Table 4).

3.2.2 Application to a microarray dataset - independent two-group data

This microarray dataset is from a study of high-risk pediatric acute lymphoblastic leukemia

(ALL), which is comprised of 191 children with ALL split into 67 minimal residual disease (MRD) positive and 124 MRD negative cases (Kang et al. 2010). The expression levels of 54675 gene probe sets were measured using the Affymetrix HG U133 Plus 2.0 platform in the pretreatment leukemia cells for each patient. The goal was to identify genes that are differentially expressed between MRD positive and negative samples. The test statistic $T$ used is the moderated $t$-statistic for independent two-groups proposed in *samr*,

$$T_{unpaired} = \frac{\overline{x}_{g2} - \overline{x}_{g1}}{s_g + s_0},$$

where $g = 1,...,G$ denotes the $g$th gene, $\overline{x}_{g1}$ and $\overline{x}_{g2}$ are respectively the sample means of the expression levels of gene $g$ of the two groups, $s_g$ is the pooled standard error of the difference between the two sample means as computed in the standard two-sample $t$-test under the assumption that the two groups have equal variance, and $s_0$ is an exchangeability factor, similar to that in Section 3.2.1, computed as a percentile from $s_g$'s, $g = 1,...,G.$ Same as in Section 3.2.1, 23 probe sets (belong to 21 unique genes) in this dataset were identified as top significantly differentially expressed using the permutation procedure in *samr* with their associated $p$-values and FDRs estimated to be 0's (Kang et al. 2010). Here we further accurately estimate the $p$-values for those probe sets using our proposed approaches, CE-Perm2 and CE-Perm2-S, and rank them. For comparisons, we also perform the crude permutations. The number of crude permutations used and the parameter settings for CE-Perm2 and CE-Perm2-S are the same as the RNA-Seq example in 3.2.1, and each procedure is repeated 100 times for each individual probe set. The results are shown in Table 5. The estimated $p$-values from all methods are similar, but our proposed CE-based algorithms achieve better performance than the crude permutation procedure in terms of the estimation precision and computational efficiency (Table 5). CE-Perm2 saves roughly 16 times of the computational time compared to crude permutation, and CE-Perm2-S further reduces the computation time to about the half of CE-Perm2 (Table 5). For comparisons, the $p$-values from the Welch's $t$-test for the 23 probe sets and the ranking of them among all genes using the Welch's $t$-test's $p$-values are presented in Supplementary Table S2.

## 4. Summary and Discussion

In summary, we develop computationally efficient algorithms for speeding up estimating small $p$-values from crude permutation tests for paired and independent two-group data through the integration of the adaptive cross-entropy method and the parameterization of the permutation sample spaces using the joint Bernoulli and the CB distributions. Simulation studies and applications to real-world gene expression datasets show that our approach achieves significant gains in computational efficiency compared with crude permutations and the existing method, SAMC. We should also note that all the examples presented involve only the comparisons of the means of the data and the test statistics used in those examples do not involve the fitting of complex models, and thus takes less amount of time to compute compared with the time of generating the permutated samples. If the test statistics used are more complicated that take more time to compute, the crude permutations and SAMC will take even longer time, since both the two procedures need much more permutated samples than our proposed approach. Additionally, we repeat each of our proposed algorithms 100 times in simulations and applications with the purpose to obtain accurate estimates of the *MCRE*s and to provide precise comparisons with crude permutations and/or SAMC. However, in practice, if the goal is only to obtain point estimates of those small p-values and rank them, it is not necessary to run such large number of parallel repetitions as 100. One can use a heuristic but practically efficient strategy discussed in (Robert and Casella 2010), which involves running a few parallel repetitions (e.g. 5-10) and checking the convergence and variations of the estimates from those repetitions. The number of repetitions can then be increased until reliable estimates with satisfactory convergence and variations are obtained. Also, note that the estimating step (Step B) in the proposed algorithms is a standard IS procedure that belongs to the general MC methods. Therefore, one can bootstrap the MC samples in a single repetition to estimate the variations, which is faster but less accurate than running a large number of parallel repetitions (Robert and Casella 2010).

Our approaches will be most useful for estimating those small to very small *p*-values (e.g. $10^{-5}$ to $10^{-15}$ as in the simulation and application examples) in permutation tests. For those not-small *p*-values, crude permutations can give accurate estimates with affordable budget and are much more straightforward to implement. Therefore, in practice, as described in the application examples, we suggest that first screening those top significant signals using crude permutations with an affordable budget or existing permutation procedures such as the one in *samr* to estimate the *p*-values or FDRs for all genes in the dataset, and identify those top significant signals using a threshold of those *p*-values or FDRs (e.g. the 41 genes in the RNA-Seq example and the 23 probe sets in the microarray example), and then apply our approaches to more accurately estimate and rank the $p$-values associated with those top signals.

As we can see in the second simulation example with the sample sizes of the two groups equal to 100, one issue with the current implementation of our approach is that the variances of the estimated $p$-values increase with the sample sizes. The underlying reason is that the number of parameters to be updated grows with sample sizes (e.g. for the simulation example with the sample sizes of the two group equal to 100, there are 100 parameters, $w_i$'s, needed to be updated in Algorithm 2), and the likelihood ratio involved in importance sampling [see Eq. (11), (14) and (A.9) in Appendix A3] becomes more and more unstable with the number of parameters growing, which has been known as the "curse of dimensionality" of the likelihood ratios when using importance sampling in high dimensional Monte Carlo simulations (Rubinstein and Glynn 2009). In that situation, either the adaptive parameter-updating step may not converge well or the proposal distribution obtained from that step can be far from the optimal proposal distribution $g^*$ (Rubinstein and Glynn 2009; Chan and Kroese 2012). With the addition of the parameter-screening step that identifies a subset of "bottleneck" parameters and then only updates that subset of parameters via the adaptive CE method, the dimensionality of the parameter space is reduced, and the aforementioned issue can be alleviated to some extent (see CE-Perm1-S and CE-Perm2-S). However, CE-Perm1-S and CE-Perm2-S still cannot handle very large sample sizes (we have tried examples with sample sizes of 500 and the adaptive parameter-updating step has issues of convergence for those cases, which are not shown here). Hence, our proposed approach is most suitable for data with sample sizes around 100 or less, such as most of differential gene expression analysis studies these days. For data with very large sample size (e.g. GWAS with more than 500 or even thousands of samples), the current implementation of our proposed approach is not suitable and methods like SAMC should be recommended. It is also worth mentioning that another approach is proposed recently to estimate small $p$-values in permutation tests (Segal et al. 2018), yet that approach is limited to special form of test statistics and not very accurate in estimating small $p$-values, and thus should be considered as an approach for "a preliminary analysis to approximate the order of magnitude of a $p$-value" (from the authors of (Segal et al. 2018)) in genomic studies.

A natural extension of this work is to stretch the current algorithms for paired and independent two-group data to multiple-group data. To that end, we need to parameterize the permutation sample space of multiple-group data by some distributions as we have done here for the paired and independent two-group permutations. One direction is to sequentially apply the CB distribution to multiple groups. For instance, if we have three groups, we can first consider the second and third groups as one single group, and then select elements for the first group by the CB distribution, and then select elements for the second group using the CB distribution again, and the remaining unselected elements are assigned to the third group. Hence, the density of the distribution that parameterizes the permutation sample space of the three-group data is the product of the densities of two CB distributions. We consider this extension as our future work.

Last but not least, despite of the popularity of permutation tests in genomic data analysis, we would like to emphasize that prior to applying our algorithms, it is crucial to carefully check the assumption of exchangeability of the data. One example of violating the exchangeability is when one is interested in testing the equality of two means of independent two-group data with unequal variances. In that case, a permutation test using the difference of two sample means can give inflated type I error rate (Huang et al. 2006). Therefore, if the data show evidence of serious violation of exchangeability, one should consider other testing procedures instead of permutation tests.

**Data availability**

The TCGA-LUSC RNA-Seq data can be downloaded from the TCGA Research Network data portal: https://portal.gdc.cancer.gov/. The ALL microarray data is available from the NCBI Gene Expression Omnibus repository under the accession number GSE11877. R programs for implementing the proposed algorithms and generating the results in the simulated and real-world datasets are available from GitHub repository under the following link:
https://github.com/shilab2017/Permutation-CE-codes.

**Acknowledgements**

We would like to thank Drs. Maureen Sartor and Xiaoquan Wen at University of Michigan for critical reading on the Methods section, and University of New Mexico Center for Advanced Research Computing and Augusta University High Performance Computing Services Core for providing the computational resources.

**Funding**

This work was partly supported by the following grants: one startup research grant from Augusta

University Medical College of Georgia (YS) and Provincial Natural Science Foundation of Sichuan Grant 2022NSFSC0666 (MW).

*Conflict of Interest:* none declared.

## Appendix

### A1. Derivation of the solution to the CE formula for paired two-group permutation test

Here we derive the solution to the CE formula for paired two-group permutation test [Eq. (14)in the main text], which is rewritten as

$$\mathbf{p}_{t+1} = \arg\max_{\mathbf{p}=[p_1,\dots,p_n]^T} D(\mathbf{p}) = \arg\max_{\mathbf{p}=[p_1,\dots,p_n]^T} \frac{1}{N}\sum_{l=1}^{N}[I\{T(\mathbf{z}_l)\geq\gamma_t\}Q(\mathbf{z}_l;\mathbf{p}_0,\mathbf{p}_t)\ln f(\mathbf{z}_l;\mathbf{p})], \tag{A.1}$$

where $Q(\mathbf{z}_l;\mathbf{p}_0,\mathbf{p}_t)=\dfrac{f(\mathbf{z};\mathbf{p}_0)}{f(\mathbf{z};\mathbf{p}_t)}$ is the likelihood ratio. Problem (A.1) can be solved by directly differentiating $D(\mathbf{p})$ with $\mathbf{p}$. Note that only the term $\ln f(\mathbf{z}_l;\mathbf{p})$ involves $\mathbf{p}$, therefore,

$$\begin{aligned}
\frac{\partial D(\mathbf{p})}{\partial p_i} &= \frac{1}{N}\sum_{l=1}^{N}[I\{T(\mathbf{z}_l)\geq\gamma_t\}Q(\mathbf{z}_l;\mathbf{p}_0,\mathbf{p}_t)\frac{\partial\ln f(\mathbf{z}_l;\mathbf{p})}{\partial p_i}] \\
&= \frac{1}{N}\sum_{l=1}^{N}[I\{T(\mathbf{z}_l)\geq\gamma_t\}Q(\mathbf{z}_l;\mathbf{p}_0,\mathbf{p}_t)\frac{\partial\ln\prod_{i=1}^{n}\{p_i^{s_{li}}(1-p_i)^{1-s_{li}}]\}}{\partial p_i}] \\
&= \frac{1}{N}\sum_{l=1}^{N}[I\{T(\mathbf{z}_l)\geq\gamma_t\}Q(\mathbf{z}_l;\mathbf{p}_0,\mathbf{p}_t)\frac{\partial\ln\prod_{i=1}^{n}\{p_i^{s_{li}}(1-p_i)^{1-s_{li}}]\}}{\partial p_i}] \\
&= \frac{1}{N}\sum_{l=1}^{N}[I\{T(\mathbf{z}_l)\geq\gamma_t\}Q(\mathbf{z}_l;\mathbf{p}_0,\mathbf{p}_t)\frac{\partial\sum_{i=1}^{n}\{s_{li}\ln p_i+(1-s_{li})\ln(1-p_i)\}}{\partial p_i}] \\
&= \frac{1}{N}\sum_{l=1}^{N}[I\{T(\mathbf{z}_l)\geq\gamma_t\}Q(\mathbf{z}_l;\mathbf{p}_0,\mathbf{p}_t)(\frac{s_{li}}{p_i}-\frac{1-s_{li}}{1-p_i})].
\end{aligned}$$

Set $\dfrac{\partial D(\mathbf{p})}{\partial p_i}$=0, we obtain the following closed form solution for $\mathbf{p}$:

$$p_i = \frac{\sum_{l=1}^{N}[I\{T(\mathbf{z}_l)\geq\gamma_t\}Q(\mathbf{z}_l;\mathbf{p}_0,\mathbf{p}_t)s_{li}]}{\sum_{l=1}^{N}[I\{T(\mathbf{z}_l)\geq\gamma_t\}Q(\mathbf{z}_l;\mathbf{p}_0,\mathbf{p}_t)]},\ \text{for } i=1,\dots,n.$$

### A2. Sampling from the CB distribution

The density of the CB distribution [Eq. (16) in the main text] is

$$f(\mathbf{d};\mathbf{w}) = \Pr(d_1,d_2,\dots,d_n \mid \textstyle\sum_{i=1}^{n} d_i = k) = \dfrac{\prod_{i=1}^{n} w_i^{d_i}}{R_k}, \tag{A.2}$$

where

$$R_k = \Pr(\sum_{i=1}^{n} d_i = k)\prod_{i=1}^{n}(1+w_i) \tag{A.3}$$

is a normalization constant and $\mathbf{w} = [w_1,...,w_n]^T$, $w_i = p_i/(1-p_i)$, $i = 1,...,n$, is the vector of odds. Under this parameterization, $w_i$'s (or equivalently, $p_i$'s ) are the parameters of the CB distribution (note that $R_k$ also involves $w_i$'s). A method called the **drafting sampling algorithm** is developed for generating random samples from the CB distribution (Chen and Liu 1997; Kroese, Taimre, and Botev 2011; Chen, Dempster, and Liu 1994). First let $R_{k-1,j}$ denote the normalization constant for the conditional distribution of $\{d_i, i \neq j\}$ given $\sum_{i \neq j} d_i = k-1$, which is defined as

$$R_{k-1,j} = \Pr\left(\sum_{i \neq j} d_i = k-1\right)\prod_{i \neq j}(1+w_i). \tag{A.4}$$

Following (Chen and Liu 1997; Kroese, Taimre, and Botev 2011), the normalization constants $R_k$ and $R_{k-1,j}$ can be recursively computed using the following relationship:

**Procedure A1:** Computation of the normalization constants of the CB distribution

Define the following quantities: $T_i = \sum_{j=1}^{n} w_j^i$ and $T_{i,j} = T_i - w_j^i$, $i = 1,...,k$, $j = 1,...,n$. Start with $R_0 = 1$ and $R_{0,j} = 1$, $j = 1,...,n$, then $R_k$ and $R_{k-1,j}$, $k = 1,...,n$, $j = 1,...,n$ can be computed as

$$R_k = \frac{1}{k}\sum_{i=1}^{k}(-1)^{i+1} T_i R_{k-i}, \tag{A.5}$$

$$R_{k-1,j} = \frac{1}{k-1}\sum_{i=1}^{k-1}(-1)^{i+1} T_{i,j} R_{k-1-i,j},\ j = 1,...,n. \tag{A.6}$$

To sample from CB distribution, we need to further define the following two quantities: the first quantity $\boldsymbol{\pi} = [\pi_1,...,\pi_n]^T$ is called the coverage probabilities of the CB distribution (Kroese, Taimre, and Botev 2011; Chen, Dempster, and Liu 1994) given as

$$\pi_j = \Pr(d_j = 1 \mid \sum_{i=1}^{n} d_i = k) = E(d_j \mid \sum_{i=1}^{n} d_i = k),\ j = 1,...,n, \tag{A.7}$$

and the second quantity $\mathbf{a} = [a_1, a_2,..., a_n]^T$ is called the coverage probability distribution given as

$$a_j = \frac{\pi_j}{k} = \frac{\Pr(d_j = 1 \mid \sum_{i=1}^{n} d_i = k)}{k},\ j = 1,...,n.$$

We can see that $\mathbf{a}$ is a normalized probability vector from $\boldsymbol{\pi}$ to form a legitimate probability distribution. The quantities $\mathbf{a}$, $\boldsymbol{\pi}$ and the normalization constants $R_k$ and $R_{k-1,j}$ have the following relationship (Kroese, Taimre, and Botev 2011; Chen, Dempster, and Liu 1994):

$$
\begin{aligned}
a_j &= \frac{\pi_j}{k} = \frac{\Pr(d_j = 1 \mid \sum_{i=1}^{n} d_i = k)}{k} \\
&= \frac{\Pr(d_j = 1, \sum_{i \neq j} d_i = k-1)}{k \Pr(\sum_{i=1}^{n} d_i = k)} \\
&= \frac{p_j R_{k-1,j} \prod_{i \neq j} (1+w_i)^{-1}}{k R_k \prod_{i=1}^{n} (1+w_i)^{-1}} \\
&= \frac{w_j R_{k-1,j}}{k R_k}, \; j = 1, \ldots, n.
\end{aligned}
\tag{A.8}
$$

The drafting sampling algorithm selects the *k* indices of 1's (recall that 1 indicates $x_i$ belongs to Group 1 and 0 indicates $x_i$ belongs to Group 2) according to $\mathbf{a}$ one by one, which is given as follows (Kroese, Taimre, and Botev 2011; Chen, Dempster, and Liu 1994):

**Procedure A2:** Sampling from CB distribution – the drafting sampling algorithm

1. Start with two sets: $S = \varnothing$ ($S$ will contain the *k* indices of 1's after the procedure and it is an empty set in this initial step) and $C = \{1, \ldots, n\}$ (which contains the current indices to be selected). Set iteration counter $t = 1$.

2. While $t \leq k$, compute $R_{k-t+1}$ and $R_{k-t,j}$, $j \in C$ with $\{w_t, t \in C\}$ based on Procedure A1, and compute the corresponding $\mathbf{a}$ based on Eq. (A.8).

3. Draw $J \sim \mathbf{a}$. Set $S = S \bigcup \{J\}$, $C = C \setminus \{J\}$ and $t = t+1$. Return to Step 2. Iterate between Step 2 and 3 until $t = k$.

4. For the *i*th index, $i = 1, \ldots, n$: if $i \in S$, then set $d_i = 1$; if $i \in C$, then set $d_i = 0$. Output $\mathbf{d} = [d_1, d_2, \ldots, d_n]^T$ as the final partition vector and determine the permutated sample $\mathbf{z}$ according to $\mathbf{d}$.

### A3. Optimization of the CE formula with the density of the CB distribution.

Plugging the density of the CB distribution (A.2) into the CE formula [Eq. (11) in the main text] gives the following optimization problem:

$$\mathbf{w}_{t+1} = \arg\max_{\mathbf{w}=[w_1,...,w_n]^T} D(\mathbf{w})$$
$$= \arg\max_{\mathbf{w}=[w_1,...,w_n]^T} \frac{1}{N}\sum_{l=1}^{N}[I\{T(\mathbf{d}_l) \geq \gamma_t\} \frac{f(\mathbf{d}_l;\mathbf{w}_0)}{f(\mathbf{d}_l;\mathbf{w}_t)} \ln f(\mathbf{d}_l;\mathbf{w})]. \quad \text{(A.9)}$$

By dropping the constant $\frac{1}{N}$ and defining $S_l := I\{T(\mathbf{d}_l) \geq \gamma_k\} \frac{f(\mathbf{d}_l;n_1,\mathbf{w}^0)}{f(\mathbf{d}_l;n_1,\mathbf{w}^k)}$ (note $S_l$ is a constant with respect to $\mathbf{w}$), problem (A.9) can be written as

$$\mathbf{w}_{k+1} = \arg\max_{\mathbf{w}=[w_1,...,w_n]^T} D(\mathbf{w}) = \arg\max_{\mathbf{w}=[w_1,...,w_n]^T} \sum_{l=1}^{N}[S_l \ln f(\mathbf{d}_l;\mathbf{w})]. \quad \text{(A.10)}$$

Further calculations by plugging $f(\mathbf{d}_l;\mathbf{w})$ (A.2) into (A.10) give

$$D(\mathbf{w}) = \sum_{l=1}^{N}[S_l \ln(\frac{\prod_{i=1}^{n} w_i^{d_{li}}}{R_{n_1}})]$$
$$= \sum_{l=1}^{N}[S_l(\sum_{i=1}^{n} d_{li} \ln w_i - \ln R_{n_1})] \quad \text{(A.11)}$$
$$= \sum_{i=1}^{n} y_i\theta_i - \ln R_{n_1} \sum_{l=1}^{N} S_l,$$

where the two new quantities $\theta_i$ and $y_i$ are defined as $\theta_i := \ln w_i$ and $y_i := \sum_{l=1}^{N} S_l d_{li}$, $i = 1,...,n$. Based on (A.11), using the new parameterization $\boldsymbol{\theta} = [\theta_1,...,\theta_n]^T$ and noting that the term $\ln R_{n_1} \sum_{l=1}^{N} S_l$ does not involve $y_i$, it can be shown that $D$ belongs to exponential family distributions and $\mathbf{y} = [y_1,...,y_n]^T$ are the sufficient statistics for $\boldsymbol{\theta}$ (Chen 2000; Bickel and Doksum 2006). Following the standard results of exponential family distributions (Bickel and Doksum 2006), the first-order derivative of $D$ is

$$\frac{\partial D}{\partial \boldsymbol{\theta}} = \mathbf{y} - E(\mathbf{y}) \quad \text{(A.12)}$$

and the maximum likelihood estimation (MLE) of the parameter $\boldsymbol{\theta}$ [equivalently, the solution to (A.10)] can be obtained by setting $\frac{\partial D}{\partial \boldsymbol{\theta}} = 0$, which is the solution to

$$\mathbf{y} = E(\mathbf{y}) = \boldsymbol{\pi} \sum_{l=1}^{N} S_l, \quad \text{(A.13)}$$

where the second equality in (A.13) follows from the definition of $\boldsymbol{\pi}$ in (A.7). Next using (A.8), (A.13) can be re-written as

$$\frac{w_i R_{k-1,i}}{R_k} = \frac{y_i}{\sum_{l=1}^{N} S_l}, \ i = 1,\dots,n. \tag{A.14}$$

In the literature, three iterative algorithms have been proposed to solve the MLE of the CB distribution, which is very similar to problem (A.14): (1) A generalized iterative scaling algorithm given in (Stern and Cover 1989). (2) An iterative proportional fitting algorithm given in (Chen, Dempster, and Liu 1994) and (Chen 2000). (3) A Newton-Raphson type algorithm given in (Chen 2000). Following (Chen 2000) and according to our empirical comparisons, the second algorithm is the most computational efficient approach. Below we provide the iterative procedure based on this the algorithm and technical details of this algorithm can be found in (Chen 2000; Chen, Dempster, and Liu 1994).

**Procedure A3:** Optimization of the CE formula with the CB density

1. Sort $\mathbf{y}$ in ascending order and let the sorted values be $\mathbf{y}' = [y'_1,\dots,y'_n]^T$.

2. Start with $w_i^{(0)} = \frac{y'_i}{\sum_{l=1}^{N} S_l}$, $i = 1,\dots,n$.

3. Subsequently update $\mathbf{w}^{(t)}$ by

$$w_i^{(t+1)} = \frac{y'_i R_{k-1,n}^{(t)}}{R_{k-1,i}^{(t)} \sum_{l=1}^{N} S_l}\Bigg|_{\mathbf{w}=\mathbf{w}^{(t)}}, i = 1,\dots,n-1; \ w_n^{(t+1)} = w_n^{(t)} = \frac{y'_i}{\sum_{l=1}^{N} S_l}$$

until convergence, where ($t$) means at the $t$th iteration.

## A4. Variance reduction with the screening method.

The idea of the screening method is that the parameter vector to be updated in the CE method often contains two types of parameters, the "bottleneck" and "non-bottleneck" parameters. The bottleneck parameters have a crucial impact on the final estimation result and their values change substantially in the CE algorithm from their initial values, while the non-bottleneck parameters have non-crucial influence on the estimation result and their values change little in the CE algorithm. The screening method identifies those bottleneck parameters according to the perturbation of the parameters updated in the CE algorithm. More details of the screening method can be found in (Rubinstein and Glynn 2009; Kroese, Taimre, and Botev 2011), and below we give the improved versions of CE-Perm1 and CE-Perm2 algorithms with one additional step of screening for the bottleneck parameters.

**Algorithm 1S:** The CE algorithm for estimating small $p$-values from paired two-group permutation test with parameter screening **(CE-Perm1-S)**

**A. Parameter-screening step:**

(1) Initialize the bottleneck parameter set $\mathbf{p}^B = [p_1,..., p_n]^T$ and non-bottleneck parameter set $\mathbf{p}^{NB} = \phi$ (an empty set), where $p_i$'s, $i = 1,...,n$ are the Bernoulli probabilities defined in Eq. (12) in the main text.

(2) Generate random samples $\mathbf{z}_1,...,\mathbf{z}_N$ from $f(\cdot;\mathbf{p}_0)$ from $n$ i.i.d. Bernoulli distributions with density defined in Eq. (13) in the main text, where $\mathbf{p}_0 = [1/2,...,1/2]^T$. Calculate the statistics $T(\mathbf{z}_1),\ldots,T(\mathbf{z}_N)$ and compute $\gamma_0$ as their $(1-\rho)$ sample quantile. Compute $\tilde{\mathbf{p}} = [\tilde{p}_1,..., \tilde{p}_n]^T$ as the solution to the following problem

$$\tilde{\mathbf{p}} = \arg\max_{\mathbf{p}=[p_1,...,p_n]^T} \frac{1}{N}\sum_{l=1}^{N}[I\{T(\mathbf{z}_l) \geq \gamma_0\} \ln f(\mathbf{z}_l;\mathbf{p})] \tag{A.15}$$

Note the only difference between Eq. (14) in the main text and (A.15) is that (A.15) does not contain the likelihood ratio $Q(\cdot)$, and hence can be solved by Eq. (15) in the main text with $Q(\cdot)$ removed.

(3) Calculate the relative perturbation for each element $\tilde{p}_i$'s, $i = 1,...,n$ as

$$\delta_i = \left| \frac{\tilde{p}_i - p_{0i}}{p_{0i}} \right|,$$

where $p_{0i}$'s, $i = 1,...,n$ are the elements of $\mathbf{p}_0$ that all equal to 1/2.

(4) If $\delta_i < \delta$, where $\delta$ is some pre-specified threshold value (e.g. 0.1), identify $p_i$ as a non-bottleneck parameter, remove it from $\mathbf{p}^B$ and put it into $\mathbf{p}^{NB}$.

(5) Repeat the above steps (2) – (4) several times until $\mathbf{p}^B$ and $\mathbf{p}^{NB}$ are stabilized or a maximum number of repetitions (e.g. 10) is reached.

**B. Adaptive parameter-updating step:**

Use the adaptive parameter-updating step described in Algorithm 1 in the main text to update only the bottleneck parameters $\mathbf{p}^B$ as identified in the above step, while the non-bottleneck parameters $\mathbf{p}^{NB}$ are fixed to the initial values (i.e. 1/2) in this step. Denote the final updated bottleneck parameters after this step as $\mathbf{p}_t^B$.

**C. Estimating step:**

The same as the estimating step described in Algorithm 1 in the main text with $\mathbf{p}_t$ as the union of $\mathbf{p}_t^B$ and $\mathbf{p}^{NB}$.

**Algorithm 2S (The CE algorithm for estimating small *p*-values from independent two-group permutation test with parameter screening – CE-Perm2-S)**

**A. Parameter-screening step:**

(1) Initialize the bottleneck parameter set $\mathbf{w}^B = [w_1,...,w_n]^T$ and non-bottleneck parameter set $\mathbf{w}^{NB} = \phi$, where $w_i$'s, $i = 1,...,n$ are the parameters of the CB distribution defined in (A.2).

(2) Generate random samples $\mathbf{z}_1,...,\mathbf{z}_N$ from the CB distribution with density $f(\cdot;\mathbf{w}_0)$ using the drafting sampling algorithm (Procedure A2), where $\mathbf{w}_0 = [1,...,1]^T$. Calculate the statistics $T(\mathbf{z}_1),\ldots,T(\mathbf{z}_N)$ and compute $\gamma_0$ as their $(1-\rho)$ sample quantile. Compute $\tilde{\mathbf{w}} = [\tilde{w}_1,...,\tilde{w}_n]^T$ as the solution to the following problem

$$\tilde{\mathbf{w}} = \underset{\mathbf{w}=[w_1,...,w_n]^T}{\arg\max} \frac{1}{N}\sum_{l=1}^{N}\left[I\{T(\mathbf{d}_l) \geq \gamma_0\} \ln f(\mathbf{d}_l;\mathbf{w})\right]. \quad \text{(A.16)}$$

Note the only difference between (A.9) and (A.16) is that (A.16) does not contain the likelihood ratio $\frac{f(\mathbf{d}_l;\mathbf{w}_0)}{f(\mathbf{d}_l;\mathbf{w}_t)}$ and hence can be solved as (A.9) using Procedure A3 with the likelihood ratio term removed.

(3) Calculate the relative perturbation for each element $\tilde{w}_i$'s, $i = 1,...,n$ as

$$\delta_i = \left|\frac{\tilde{w}_i - w_{0i}}{w_{0i}}\right|,$$

where $w_{0i}$'s, $i = 1,...,n$ are the elements of $\mathbf{w}_0$ that all equal to ones.

(4) If $\delta_i < \delta$, where $\delta$ is some pre-specified threshold value (e.g. 0.1), identify $w_i$ as a non-bottleneck parameter, remove it from $\mathbf{w}^B$ and put it into $\mathbf{w}^{NB}$.

(4) Repeat the above steps (2) - (4) several times until $\mathbf{w}^B$ and $\mathbf{w}^{NB}$ are stabilized or a maximum number of repetitions (e.g. 10) is reached.

**B. Adaptive parameter-updating step:**

Use the adaptive parameter-updating step described in Algorithm 2 in the main text to update only the bottleneck parameters $\mathbf{w}^B$ as identified in the above step, while the non-bottleneck parameters $\mathbf{w}^{NB}$ are fixed to the initial values (i.e. 1) in this step. Denote the final updated bottleneck parameters after this step as $\mathbf{w}_t^B$.

**C. Estimating step:**

The same as the estimating step described in Algorithm 2 in the main text with $\mathbf{w}_t$ as the union of $\mathbf{w}_t^B$ and $\mathbf{w}^{NB}$.

## Tables

**Table 1.** Results from the simulated example of paired two-group permutation test

| **Order of magnitude of *P*** | ***t*-test** | **Crude** | | **CE-Perm1** | | **CE-Perm1-S** | |
|---|---|---|---|---|---|---|---|
| | | ***P*** **(*MCRE*)** | ***N*** **(Time)** | ***P*** **(*MCRE*)** | ***N*** **(Time)** | ***P*** **(*MCRE*)** | ***N*** **(Time)** |
| $10^{-5}$ | $4.42\times10^{-5}$ | $4.47\times10^{-5}$<br>($4.19\times10^{-3}$) | $10^{7}$<br>($7.32\times10^{3}$) | $4.49\times10^{-5}$<br>($2.57\times10^{-3}$) | $1.63\times10^{4}$<br>(75.61) | $4.43\times10^{-5}$<br>($1.97\times10^{-3}$) | $1.62\times10^{4}$<br>(73.07) |
| $10^{-6}$ | $1.58\times10^{-6}$ | $1.63\times10^{-6}$<br>($8.10\times10^{-3}$) | $10^{8}$<br>($7.24\times10^{4}$) | $1.62\times10^{-6}$<br>($3.85\times10^{-3}$) | $1.74\times10^{4}$<br>(86.42) | $1.66\times10^{-6}$<br>($2.12\times10^{-3}$) | $1.72\times10^{4}$<br>(80.93) |
| $10^{-7}$ | $1.44\times10^{-7}$ | $1.55\times10^{-7}$<br>($8.73\times10^{-3}$) | $10^{9}$<br>($7.15\times10^{5}$) | $1.57\times10^{-7}$<br>($4.24\times10^{-3}$) | $1.80\times10^{4}$<br>(87.92) | $1.59\times10^{-7}$<br>($2.59\times10^{-3}$) | $1.81\times10^{4}$<br>(82.54) |
| $10^{-8}$ | $1.18\times10^{-8}$ | $1.24\times10^{-8}$<br>($9.00\times10^{-3}$) | $10^{10}$<br>($6.98 \times10^{6}$) | $1.28\times10^{-8}$<br>($5.13\times10^{-3}$) | $1.92\times10^{4}$<br>(92.03) | $1.22\times10^{-8}$<br>($3.97\times10^{-3}$) | $1.91\times10^{4}$<br>(88.17) |
| $10^{-9}$ | $3.26\times10^{-9}$ | - | - | $3.65\times10^{-9}$<br>($6.92\times10^{-3}$) | $1.99\times10^{4}$<br>(105.21) | $3.56\times10^{-9}$<br>($4.51\times10^{-3}$) | $1.97\times10^{4}$<br>(97.36) |
| $10^{-10}$ | $2.35\times10^{-10}$ | - | - | $3.15\times10^{-10}$<br>($9.64\times10^{-3}$) | $2.02\times10^{4}$<br>(106.42) | $3.11\times10^{-10}$<br>($5.68\times10^{-3}$) | $2.00\times10^{4}$<br>(101.95) |
| $10^{-11}$ | $6.17\times10^{-11}$ | - | - | $8.25\times10^{-11}$<br>($1.22\times10^{-2}$) | $2.14\times10^{4}$<br>(110.38) | $8.32\times10^{-11}$<br>($6.25\times10^{-3}$) | $2.15\times10^{4}$<br>(104.59) |
| $10^{-12}$ | $4.08\times10^{-12}$ | - | - | $6.09\times10^{-12}$<br>($1.63\times10^{-2}$) | $2..18\times10^{4}$<br>(114.42) | $6.03\times10^{-12}$<br>($7.61\times10^{-3}$) | $2.19\times10^{4}$<br>(108.37) |
| $10^{-13}$ | $2.60\times10^{-13}$ | - | - | $4.73\times10^{-13}$<br>($1.97\times10^{-2}$) | $2.32\times10^{4}$<br>(120.03) | $4.65\times10^{-13}$<br>($8.82\times10^{-3}$) | $2.33\times10^{4}$<br>(115.62) |
| $10^{-14}$ | $1.61\times10^{-14}$ | - | - | $4.83\times10^{-14}$<br>($2.11\times10^{-2}$) | $2.39\times10^{4}$<br>(132.38) | $4.77\times10^{-14}$<br>($9.27\times10^{-3}$) | $2.38\times10^{4}$<br>(121.74) |
| $10^{-15}$ | $3.99\times10^{-15}$ | - | - | $9.98\times10^{-15}$<br>($2.37\times10^{-2}$) | $2.43\times10^{4}$<br>(141.97) | $9.91\times10^{-15}$<br>($1.12\times10^{-2}$) | $2.42\times10^{4}$<br>(129.26) |

***t*-test:** the *p*-value from one-sample *t*-test that gives the order of the *p*-value from permutation test, but note that it is different from the permutation; ***P*:** the final estimated *p*-value from each procedure, which is the average of the 100 runs; ***MCRE*:** the Monte Carlo relative error as defined in the main text; ***N*:** the number of permutations or random samples used in a single run of each procedure. For CE-Perm1 and CE-Perm1-S, *N* is reported as the average of the 100 runs since the adaptive parameter-updating steps may take different numbers of iterations to converge for different runs; **Time:** total CPU time in seconds on the AMD Opteron 6272, 2.1 GHz CPU for the 100 runs.

**Table 2.** Results from the first simulated example of unpaired two-group permutation test with equal sample sizes of 20

| **Order of magnitude of *P*** | ***t*-test** | **Crude** | | **SAMC** | | **CE-Perm2** | | **CE-Perm2-S** | |
|---|---|---|---|---|---|---|---|---|---|
| | | ***P*** **(*MCRE*)** | ***N*** **(Time)** | ***P*** **(*MCRE*)** | ***N*** **(Time)** | ***P*** **(*MCRE*)** | ***N*** **(Time)** | ***P*** **(*MCRE*)** | ***N*** **(Time)** |
| $10^{-5}$ | $6.89\times10^{-5}$ | $6.72\times10^{-5}$ ($3.44\times10^{-3}$) | $10^{7}$ ($2.29\times10^{4}$) | $6.62\times10^{-5}$ ($5.03\times10^{-3}$) | $1.2\times10^{6}$ ($6.72\times10^{4}$) | $6.67\times10^{-5}$ ($2.08\times10^{-3}$) | $1.65\times10^{4}$ ($9.04\times10^{2}$) | $6.70\times10^{-5}$ ($2.13\times10^{-3}$) | $1.63\times10^{4}$ ($1.24\times10^{3}$) |
| $10^{-6}$ | $4.49\times10^{-6}$ | $4.77\times10^{-6}$ ($4.72\times10^{-3}$) | $10^{8}$ ($2.12\times10^{5}$) | $4.66\times10^{-6}$ ($4.21\times10^{-3}$) | $1.2\times10^{6}$ ($6.82\times10^{4}$) | $4.71\times10^{-6}$ ($2.32\times10^{-3}$) | $1.76\times10^{4}$ ($2.43\times10^{3}$) | $4.69\times10^{-6}$ ($2.04\times10^{-3}$) | $1.76\times10^{4}$ ($1.87\times10^{3}$) |
| $10^{-7}$ | $2.76\times10^{-7}$ | $3.68\times10^{-7}$ ($5.21\times10^{-3}$) | $10^{9}$ ($2.17\times10^{6}$) | $3.63\times10^{-7}$ ($5.57\times10^{-3}$) | $1.2\times10^{6}$ ($6.97\times10^{4}$) | $3.68\times10^{-7}$ ($2.34\times10^{-3}$) | $1.84\times10^{4}$ ($2.44\times10^{3}$) | $3.64\times10^{-7}$ ($2.21\times10^{-3}$) | $1.85\times10^{4}$ ($2.16\times10^{3}$) |
| $10^{-8}$ | $1.71\times10^{-8}$ | $2.81\times10^{-8}$ ($5.97\times10^{-3}$) | $10^{10}$ ($2.09\times10^{7}$) | $2.88\times10^{-8}$ ($6.47\times10^{-3}$) | $1.2\times10^{6}$ ($7.03\times10^{4}$) | $2.82\times10^{-8}$ ($2.87\times10^{-3}$) | $1.88\times10^{4}$ ($2.67\times10^{3}$) | $2.86\times10^{-8}$ ($2.49\times10^{-3}$) | $1.90\times10^{4}$ ($2.38\times10^{3}$) |
| $10^{-9}$ | $1.11\times10^{-9}$ | - | - | $2.37\times10^{-9}$ ($7.23\times10^{-3}$) | $1.2\times10^{6}$ ($7.12\times10^{4}$) | $2.48\times10^{-9}$ ($3.45\times10^{-3}$) | $1.98\times10^{4}$ ($2.91\times10^{3}$) | $2.43\times10^{-9}$ ($2.72\times10^{-3}$) | $1.99\times10^{4}$ ($2.64\times10^{3}$) |
| $10^{-10}$ | $3.82\times10^{-10}$ | - | - | $4.98\times10^{-10}$ ($8.64\times10^{-3}$) | $1.2\times10^{6}$ ($7.19\times10^{4}$) | $4.86\times10^{-10}$ ($4.01\times10^{-3}$) | $2.06\times10^{4}$ ($3.06\times10^{3}$) | $4.95\times10^{-10}$ ($3.14\times10^{-3}$) | $2.07\times10^{4}$ ($2.92\times10^{3}$) |
| $10^{-11}$ | $7.86\times10^{-11}$ | - | - | $9.06\times10^{-11}$ ($9.21\times10^{-3}$) | $1.2\times10^{6}$ ($7.24\times10^{4}$) | $9.21\times10^{-11}$ ($4.83\times10^{-3}$) | $2.13\times10^{4}$ ($3.17\times10^{3}$) | $9.13\times10^{-11}$ ($2.89\times10^{-3}$) | $2.12\times10^{4}$ ($2.97\times10^{3}$) |
| $10^{-12}$ | $6.09\times10^{-12}$ | - | - | $7.52\times10^{-12}$ ($9.86\times10^{-2}$) | $1.2\times10^{6}$ ($7.41\times10^{4}$) | $7.44\times10^{-12}$ ($5.65\times10^{-3}$) | $2.26\times10^{4}$ ($3.43\times10^{3}$) | $7.38\times10^{-12}$ ($3.51\times10^{-3}$) | $2.24\times10^{4}$ ($3.03\times10^{3}$) |
| $10^{-13}$ | $5.21\times10^{-13}$ | - | - | $6.82\times10^{-13}$ ($1.16\times10^{-2}$) | $1.2\times10^{6}$ ($7.53\times10^{4}$) | $6.90\times10^{-13}$ ($6.23\times10^{-3}$) | $2.37\times10^{4}$ ($3.69\times10^{3}$) | $6.85\times10^{-13}$ ($3.35\times10^{-3}$) | $2.39\times10^{4}$ ($3.21\times10^{3}$) |
| $10^{-14}$ | $4.94\times10^{-14}$ | - | - | $7.13\times10^{-14}$ ($1.34\times10^{-2}$) | $1.2\times10^{6}$ ($7.67\times10^{4}$) | $7.26\times10^{-14}$ ($7.14\times10^{-3}$) | $2.46\times10^{4}$ ($3.86\times10^{3}$) | $7.29\times10^{-14}$ ($4.21\times10^{-3}$) | $2.45\times10^{4}$ ($3.43\times10^{3}$) |
| $10^{-15}$ | $5.18\times10^{-15}$ | - | - | $7.51\times10^{-15}$ ($1.47\times10^{-2}$) | $1.2\times10^{6}$ ($7.89\times10^{4}$) | $7.36\times10^{-15}$ ($7.68\times10^{-3}$) | $2.55\times10^{4}$ ($4.11\times10^{3}$) | $7.44\times10^{-15}$ ($4.73\times10^{-3}$) | $2.57\times10^{4}$ ($3.57\times10^{3}$) |

The meanings of the column names are the same as Table 1.

**Table 3.** Results from the second simulated example of unpaired two-group permutation test with equal sample sizes of 100

| **Order of magnitude of *P*** | ***t*-test** | **SAMC I** | | **SAMC II** | | **CE-Perm2** | | **CE-Perm2-S** | |
|---|---|---|---|---|---|---|---|---|---|
| | | ***P*** **(*MCRE*)** | ***N*** **(Time)** | ***P*** **(*MCRE*)** | ***N*** **(Time)** | ***P*** **(*MCRE*)** | ***N*** **(Time)** | ***P*** **(*MCRE*)** | ***N*** **(Time)** |
| $10^{-5}$ | $2.34\times10^{-5}$ | $2.57\times10^{-5}$ ($2.18\times10^{-3}$) | $5.2\times10^{6}$ ($2.76\times10^{5}$) | $2.68\times10^{-5}$ ($5.63\times10^{-3}$) | $1.2\times10^{6}$ ($7.66\times10^{4}$) | $2.52\times10^{-5}$ ($5.11\times10^{-3}$) | $3.16\times10^{4}$ ($2.37\times10^{4}$) | $2.49\times10^{-5}$ ($2.29\times10^{-3}$) | $3.13\times10^{4}$ ($1.06\times10^{4}$) |
| $10^{-6}$ | $4.80\times10^{-6}$ | $5.12\times10^{-6}$ ($2.52\times10^{-3}$) | $5.2\times10^{6}$ ($2.87\times10^{5}$) | $5.18\times10^{-6}$ ($5.74\times10^{-3}$) | $1.2\times10^{6}$ ($7.72\times10^{4}$) | $5.23\times10^{-6}$ ($5.53\times10^{-3}$) | $3.28\times10^{4}$ ($2.56\times10^{4}$) | $5.26\times10^{-6}$ ($2.67\times10^{-3}$) | $3.29\times10^{4}$ ($1.21\times10^{4}$) |
| $10^{-7}$ | $8.91\times10^{-7}$ | $9.58\times10^{-7}$ ($2.71\times10^{-3}$) | $5.2\times10^{6}$ ($3.03\times10^{5}$) | $9.45\times10^{-7}$ ($5.93\times10^{-3}$) | $1.2\times10^{6}$ ($7.80\times10^{4}$) | $9.54\times10^{-7}$ ($6.87\times10^{-3}$) | $3.41\times10^{4}$ ($2.71\times10^{4}$) | $9.67\times10^{-7}$ ($2.94\times10^{-3}$) | $3.42\times10^{4}$ ($1.36\times10^{4}$) |
| $10^{-8}$ | $2.31\times10^{-8}$ | $2.65\times10^{-8}$ ($3.16\times10^{-3}$) | $5.2\times10^{6}$ ($3.18\times10^{5}$) | $2.60\times10^{-8}$ ($6.58\times10^{-3}$) | $1.2\times10^{6}$ ($8.02\times10^{4}$) | $2.62\times10^{-8}$ ($8.67\times10^{-3}$) | $3.50\times10^{4}$ ($3.34\times10^{4}$) | $2.57\times10^{-8}$ ($3.51\times10^{-3}$) | $3.52\times10^{4}$ ($1.58\times10^{4}$) |
| $10^{-9}$ | $3.27\times10^{-9}$ | $3.89\times10^{-9}$ ($3.57\times10^{-3}$) | $5.2\times10^{6}$ ($3.43\times10^{5}$) | $3.82\times10^{-9}$ ($7.96\times10^{-3}$) | $1.2\times10^{6}$ ($8.07\times10^{4}$) | $3.64\times10^{-9}$ ($1.23\times10^{-2}$) | $3.65\times10^{4}$ ($4.16\times10^{4}$) | $3.69\times10^{-9}$ ($4.13\times10^{-3}$) | $3.64\times10^{4}$ ($1.97\times10^{4}$) |
| $10^{-10}$ | $4.27\times10^{-10}$ | $5.41\times10^{-10}$ ($6.86\times10^{-3}$) | $5.2\times10^{6}$ ($3.69\times10^{5}$) | $5.19\times10^{-10}$ ($1.21\times10^{-2}$) | $1.2\times10^{6}$ ($8.14\times10^{4}$) | $5.13\times10^{-10}$ ($1.67\times10^{-2}$) | $3.82\times10^{4}$ ($4.18\times10^{4}$) | $5.28\times10^{-10}$ ($6.75\times10^{-3}$) | $3.80\times10^{4}$ ($2.26\times10^{4}$) |
| $10^{-11}$ | $5.18\times10^{-11}$ | $5.77\times10^{-11}$ ($9.17\times10^{-3}$) | $5.2\times10^{6}$ ($3.87\times10^{5}$) | $5.81\times10^{-11}$ ($1.82\times10^{-2}$) | $1.2\times10^{6}$ ($8.27\times10^{4}$) | $5.94\times10^{-11}$ ($2.17\times10^{-2}$) | $4.04\times10^{4}$ ($4.76\times10^{4}$) | $5.83\times10^{-11}$ ($9.28\times10^{-3}$) | $4.01\times10^{4}$ ($2.68\times10^{4}$) |
| $10^{-12}$ | $5.86\times10^{-12}$ | $6.92\times10^{-12}$ ($1.35\times10^{-2}$) | $5.2\times10^{6}$ ($4.21\times10^{5}$) | $6.98\times10^{-12}$ ($2.37\times10^{-2}$) | $1.2\times10^{6}$ ($8.53\times10^{4}$) | $6.73\times10^{-12}$ ($2.76\times10^{-2}$) | $4.18\times10^{4}$ ($5.21\times10^{4}$) | $6.62\times10^{-12}$ ($1.17\times10^{-2}$) | $4.21\times10^{4}$ ($2.95\times10^{4}$) |
| $10^{-13}$ | $6.21\times10^{-13}$ | $7.79\times10^{-13}$ ($1.62\times10^{-2}$) | $5.2\times10^{6}$ ($4.58\times10^{5}$) | $7.71\times10^{-13}$ ($2.92\times10^{-2}$) | $1.2\times10^{6}$ ($8.79\times10^{4}$) | $7.64\times10^{-13}$ ($3.13\times10^{-2}$) | $4.34\times10^{4}$ ($5.69\times10^{4}$) | $7.53\times10^{-13}$ ($1.32\times10^{-2}$) | $4.36\times10^{4}$ ($3.32\times10^{4}$) |
| $10^{-14}$ | $6.21\times10^{-14}$ | $7.98\times10^{-14}$ ($1.89\times10^{-2}$) | $5.2\times10^{6}$ ($4.89\times10^{5}$) | $7.90\times10^{-14}$ ($3.53\times10^{-2}$) | $1.2\times10^{6}$ ($9.04\times10^{4}$) | $8.07\times10^{-14}$ ($3.62\times10^{-2}$) | $4.62\times10^{4}$ ($6.25\times10^{4}$) | $7.96\times10^{-14}$ ($1.46\times10^{-2}$) | $4.59\times10^{4}$ ($3.73\times10^{4}$) |
| $10^{-15}$ | $5.89\times10^{-15}$ | $7.73\times10^{-15}$ ($2.07\times10^{-2}$) | $5.2\times10^{6}$ ($5.26\times10^{5}$) | $7.68\times10^{-15}$ ($3.87\times10^{-2}$) | $1.2\times10^{6}$ ($9.37\times10^{4}$) | $7.52\times10^{-15}$ ($4.15\times10^{-2}$) | $4.78\times10^{4}$ ($6.88\times10^{4}$) | $7.64\times10^{-15}$ ($1.69\times10^{-2}$) | $4.82\times10^{4}$ ($4.14\times10^{4}$) |

The meanings of the column names are the same as Table 1.

**Table 4.** The results for the top 10 significant genes of the TCGA-LUSC RNA-Seq data.

| Gene | CE-Perm1 | | CE-Perm1-S | |
|---|---|---|---|---|
| | ***P*-value** | ***MCRE*** | ***P*-value** | ***MCRE*** |
| SDHC | $5.26\times10^{-15}$ | $3.39\times10^{-02}$ | $5.43\times10^{-15}$ | $1.37\times10^{-02}$ |
| FRMD4B | $6.94\times10^{-15}$ | $3.22\times10^{-02}$ | $7.13\times10^{-15}$ | $2.28\times10^{-02}$ |
| C20orf135 | $8.15\times10^{-15}$ | $3.34\times10^{-02}$ | $8.08\times10^{-15}$ | $1.94\times10^{-02}$ |
| CBWD3 | $8.42\times10^{-15}$ | $3.17\times10^{-02}$ | $8.57\times10^{-15}$ | $2.15\times10^{-02}$ |
| GSC2 | $8.60\times10^{-15}$ | $3.46\times10^{-02}$ | $8.89\times10^{-15}$ | $1.75\times10^{-02}$ |
| NCRNA00085 | $4.85\times10^{-14}$ | $2.97\times10^{-02}$ | $4.80\times10^{-14}$ | $2.17\times10^{-02}$ |
| CACNG4 | $5.16\times10^{-14}$ | $3.04\times10^{-02}$ | $5.30\times10^{-14}$ | $1.36\times10^{-02}$ |
| C8orf41 | $8.21\times10^{-14}$ | $3.11\times10^{-02}$ | $8.28\times10^{-14}$ | $2.40\times10^{-02}$ |
| FAM19A3 | $8.46\times10^{-14}$ | $2.87\times10^{-02}$ | $8.42\times10^{-14}$ | $1.51\times10^{-02}$ |
| DTX3L | $9.72\times10^{-14}$ | $3.15\times10^{-02}$ | $1.02\times10^{-13}$ | $1.34\times10^{-02}$ |

See Supplementary Table S1 for the detailed results. The Total CPU time of each procedure is, Crude: $2.61\times10^{6}$ s, CE-Perm1: $4.37\times10^{3}$ s, CE-Perm1-S: $3.97\times10^{3}$ s, on an AMD Opteron 6272, 2.1 GHz CPU.

**Table 5.** The estimated exact *p*-values for the top 23 significant probe sets of the MRD microarray data.

Total CPU time of each procedure: Crude - $9.85\times10^{6}$ s, CE-Perm2 - $6.26\times10^{5}$ s, CE-Perm2-S - $2.83\times10^{5}$ s.

| **Probe Set ID** | **Gene Symbol** | **Crude *P*-value (*MCRE*)** | **CE-Perm2 *P*-value (*MCRE*)** | **CE-Perm2-S *P*-value (*MCRE*)** | **Description** |
|---|---|---|---|---|---|
| 242747_at | --- | $2.40\times10^{-9}$ ($2.72\times10^{-1}$) | $2.71\times10^{-9}$ ($2.91\times10^{-2}$) | $2.58\times10^{-9}$ ($1.37\times10^{-2}$) | NCI_CGAP_Brn35 Homo sapiens cDNA clone IMAGE:2616532 3' mRNA sequence |
| 1564310_a_at | PARP15 | $3.80\times10^{-9}$ ($2.08\times10^{-1}$) | $4.39\times10^{-9}$ ($1.94\times10^{-2}$) | $4.31\times10^{-9}$ ($1.02\times10^{-2}$) | poly (ADP-ribose) polymerase family, member 15 |
| 201718_s_at | EPB41L2 | $3.20\times10^{-9}$ ($2.47\times10^{-1}$) | $4.41\times10^{-9}$ ($1.65\times10^{-2}$) | $4.17\times10^{-9}$ ($9.61\times10^{-3}$) | erythrocyte membrane protein band 4.1-like 2 |
| 219032_x_at | OPN3 | $3.02\times10^{-8}$ ($8.54\times10^{-2}$) | $2.89\times10^{-8}$ ($3.19\times10^{-2}$) | $2.83\times10^{-8}$ ($1.47\times10^{-2}$) | opsin 3 |
| 201719_s_at | EPB41L2 | $6.82\times10^{-8}$ ($4.34\times10^{-2}$) | $7.16\times10^{-8}$ ($1.22\times10^{-2}$) | $7.19\times10^{-8}$ ($8.33\times10^{-3}$) | erythrocyte membrane protein band 4.1-like 2 |
| 205429_s_at | MPP6 | $8.98\times10^{-8}$ ($4.52\times10^{-2}$) | $8.67\times10^{-8}$ ($5.78\times10^{-3}$) | $8.74\times10^{-8}$ ($4.76\times10^{-3}$) | membrane protein, palmitoylated 6 (MAGUK p55 subfamily member 6) |
| 1553380_at | PARP15 | $1.12\times10^{-7}$ ($4.62\times10^{-2}$) | $1.07\times10^{-7}$ ($9.81\times10^{-3}$) | $1.03\times10^{-7}$ ($6.85\times10^{-3}$) | poly (ADP-ribose) polymerase family, member 15 |
| 207426_s_at | TNFSF4 | $1.65\times10^{-7}$ ($3.67\times10^{-2}$) | $1.58\times10^{-7}$ ($1.65\times10^{-2}$) | $1.52\times10^{-7}$ ($8.76\times10^{-3}$) | tumor necrosis factor (ligand) superfamily, member 4 (tax-transcriptionally activated glycoprotein 1, 34kDa) |
| 209286_at | CDC42EP3 | $1.76\times10^{-7}$ ($3.49\times10^{-2}$) | $1.73\times10^{-7}$ ($1.46\times10^{-2}$) | $1.78\times10^{-7}$ ($7.34\times10^{-3}$) | CDC42 effector protein (Rho GTPase binding) 3 |
| 221841_s_at | KLF4 | $2.14\times10^{-7}$ ($2.84\times10^{-2}$) | $2.00\times10^{-7}$ ($9.05\times10^{-3}$) | $2.08\times10^{-7}$ ($6.57\times10^{-3}$) | Kruppel-like factor 4 (gut) |
| 227336_at | DTX1 | $4.17\times10^{-7}$ ($2.12\times10^{-2}$) | $4.27\times10^{-7}$ ($5.81\times10^{-3}$) | $4.22\times10^{-7}$ ($5.02\times10^{-3}$) | deltex homolog 1 (Drosophila) |
| 225685_at | --- | $4.75\times10^{-7}$ ($2.07\times10^{-2}$) | $4.89\times10^{-7}$ ($6.18\times10^{-3}$) | $4.78\times10^{-7}$ ($4.39\times10^{-3}$) | CDNA FLJ31353 fis, clone MESAN2000264 |
| 213358_at | KIAA0802 | $6.30\times10^{-7}$ ($1.75\times10^{-2}$) | $6.16\times10^{-7}$ ($7.66\times10^{-3}$) | $6.14\times10^{-7}$ ($4.27\times10^{-3}$) | KIAA0802 |
| 219990_at | E2F8 | $6.57\times10^{-7}$ ($1.60\times10^{-2}$) | $6.60\times10^{-7}$ ($9.52\times10^{-3}$) | $6.63\times10^{-7}$ ($3.59\times10^{-3}$) | E2F transcription factor 8 |
| 204562_at | IRF4 | $6.78\times10^{-7}$ ($1.76\times10^{-2}$) | $6.70\times10^{-7}$ ($5.97\times10^{-3}$) | $6.65\times10^{-7}$ ($4.38\times10^{-3}$) | interferon regulatory factor 4 |
| 213817_at | --- | $8.91\times10^{-7}$ ($1.48\times10^{-2}$) | $8.71\times10^{-7}$ ($5.63\times10^{-3}$) | $8.79\times10^{-7}$ ($3.04\times10^{-3}$) | CDNA FLJ13601 fis, clone PLACE1010069 |
| 201710_at | MYBL2 | $8.89\times10^{-7}$ ($1.44\times10^{-2}$) | $8.95\times10^{-7}$ ($5.53\times10^{-3}$) | $8.93\times10^{-7}$ ($2.26\times10^{-3}$) | v-myb myeloblastosis viral oncogene homolog (avian)-like 2 |
| 232539_at | --- | $9.79\times10^{-7}$ ($1.38\times10^{-2}$) | $9.58\times10^{-7}$ ($6.01\times10^{-3}$) | $9.62\times10^{-7}$ ($2.89\times10^{-3}$) | MRNA; cDNA DKFZp761H1023 (from clone DKFZp761H1023) |
| 218589_at | P2RY5 | $1.36\times10^{-6}$ ($1.23\times10^{-2}$) | $1.37\times10^{-6}$ ($5.15\times10^{-3}$) | $1.35\times10^{-6}$ ($2.63\times10^{-3}$) | purinergic receptor P2Y, G-protein coupled, 5 |
| 218899_s_at | BAALC | $1.54\times10^{-6}$ ($1.28\times10^{-2}$) | $1.57\times10^{-6}$ ($4.29\times10^{-3}$) | $1.55\times10^{-6}$ ($2.24\times10^{-3}$) | brain and acute leukemia, cytoplasmic |
| 225688_s_at | PHLDB2 | $2.04\times10^{-6}$ ($9.12\times10^{-3}$) | $2.06\times10^{-6}$ ($6.36\times10^{-3}$) | $2.05\times10^{-6}$ ($2.19\times10^{-3}$) | pleckstrin homology-like domain, family B, member 2 |
| 242051_at | CD99 | $5.66\times10^{-6}$ ($5.58\times10^{-3}$) | $5.66\times10^{-6}$ ($5.02\times10^{-3}$) | $5.67\times10^{-6}$ ($2.87\times10^{-3}$) | CD99 molecule |
| 220448_at | KCNK12 | $7.03\times10^{-6}$ ($5.31\times10^{-3}$) | $7.08\times10^{-6}$ ($4.79\times10^{-3}$)* | $7.07\times10^{-6}$ ($2.53\times10^{-3}$) | potassium channel, subfamily K, member 12 |

*For this probe set, there is one single run of CE-Perm2 where the adaptive updating step does not converge. The *p*-value and *MCRE* presented are based on the converged 99 runs.

**Supplementary Table S1.** The results for the top 41 significant genes of the TCGA-LUSC RNA-Seq data.

| Gene | Description | P (crude) | SD (crude) | MCRE (crude) | P (CE-Perm1) | SD (CE-Perm1) | MCRE (CE-Perm1) | P (CE-Perm1-S) | SD (CE-Perm1-S) | MCRE (CE-Perm1-S) |
|---|---|---|---|---|---|---|---|---|---|---|
| SDHC | succinate dehydrogenase complex subunit C | 0 | 0 | NA | 5.26E-15 | 1.75E-15 | 3.39E-02 | 5.43E-15 | 7.44E-16 | 1.37E-02 |
| FRMD4B | FERM domain containing 4B | 0 | 0 | NA | 6.94E-15 | 2.19E-15 | 3.22E-02 | 7.13E-15 | 1.63E-15 | 2.28E-02 |
| C20orf135 | abhydrolase domain containing 16B | 0 | 0 | NA | 8.15E-15 | 2.69E-15 | 3.34E-02 | 8.08E-15 | 1.65E-15 | 1.94E-02 |
| CBWD3 | COBW domain containing 3 | 0 | 0 | NA | 8.42E-15 | 2.67E-15 | 3.17E-02 | 8.57E-15 | 1.84E-15 | 2.15E-02 |
| GSC2 | goosecoid homeobox 2 | 0 | 0 | NA | 8.60E-15 | 2.98E-15 | 3.46E-02 | 8.89E-15 | 1.56E-15 | 1.75E-02 |
| NCRNA00085 | sperm acrosome associated 6 | 0 | 0 | NA | 4.85E-14 | 1.44E-14 | 2.97E-02 | 4.80E-14 | 1.04E-14 | 2.17E-02 |
| CACNG4 | calcium voltage-gated channel auxiliary subunit gamma 4 | 0 | 0 | NA | 5.16E-14 | 1.53E-14 | 3.04E-02 | 5.30E-14 | 7.21E-15 | 1.36E-02 |
| C8orf41 | TELO2 interacting protein 2 | 0 | 0 | NA | 8.21E-14 | 2.55E-14 | 3.11E-02 | 8.28E-14 | 2.02E-14 | 2.40E-02 |
| FAM19A3 | family with sequence similarity 19 member A3, C-C motif chemokine like | 0 | 0 | NA | 8.46E-14 | 2.43E-14 | 2.87E-02 | 8.42E-14 | 1.29E-14 | 1.51E-02 |
| DTX3L | deltex E3 ubiquitin ligase 3L | 0 | 0 | NA | 9.72E-14 | 3.06E-14 | 3.15E-02 | 1.02E-13 | 1.37E-14 | 1.34E-02 |
| ARHGEF35 | Rho guanine nucleotide exchange factor 35 | 0 | 0 | NA | 1.14E-13 | 3.21E-14 | 2.82E-02 | 1.10E-13 | 1.90E-14 | 1.73E-02 |
| PRAMEF1 | PRAME family member 1 | 0 | 0 | NA | 1.68E-13 | 4.57E-14 | 2.72E-02 | 1.62E-13 | 2.58E-14 | 1.59E-02 |
| SLC26A2 | solute carrier family 26 member 2 | 0 | 0 | NA | 2.69E-13 | 5.89E-14 | 2.19E-02 | 2.55E-13 | 3.01E-14 | 1.18E-02 |
| MPO | myeloperoxidase | 0 | 0 | NA | 3.32E-13 | 6.14E-14 | 1.85E-02 | 3.25E-13 | 3.48E-14 | 1.07E-02 |
| SLC17A4 | solute carrier family 17 member 4 | 0 | 0 | NA | 7.81E-13 | 2.12E-13 | 2.53E-02 | 7.92E-13 | 9.98E-14 | 1.26E-02 |
| RNF222 | ring finger protein 222 | 0 | 0 | NA | 8.73E-13 | 1.41E-13 | 1.66E-02 | 8.78E-13 | 1.00E-13 | 1.14E-02 |
| DDO | D-aspartate oxidase | 0 | 0 | NA | 1.37E-12 | 3.81E-13 | 2.78E-02 | 1.38E-12 | 2.53E-13 | 1.83E-02 |
| ADCK4 | coenzyme Q8B | 0 | 0 | NA | 6.46E-12 | 1.55E-12 | 2.47E-02 | 6.60E-12 | 1.11E-12 | 1.68E-02 |
| PFKFB4 | 6-phosphofructo-2-kinase/fructose-2,6-biphosphatase 4 | 0 | 0 | NA | 7.09E-12 | 1.57E-12 | 2.21E-02 | 6.98E-12 | 7.61E-13 | 1.09E-02 |
| APAF1 | apoptotic peptidase activating factor 1 | 0 | 0 | NA | 3.60E-11 | 6.84E-12 | 1.90E-02 | 3.52E-11 | 4.29E-12 | 1.22E-02 |
| LOC121952 | methyltransferase like 21E, pseudogene | 0 | 0 | NA | 4.54E-11 | 6.99E-12 | 1.54E-02 | 4.40E-11 | 2.05E-12 | 4.67E-03 |
| PRRG3 | proline rich and Gla domain 3 | 0 | 0 | NA | 5.41E-11 | 5.52E-12 | 1.02E-02 | 5.33E-11 | 2.51E-12 | 4.71E-03 |
| LOC389634 | long intergenic non-protein coding RNA 937 | 0 | 0 | NA | 6.72E-11 | 9.95E-12 | 1.48E-02 | 6.83E-11 | 7.38E-12 | 1.08E-02 |
| C6orf132 | chromosome 6 open reading frame 132 | 0 | 0 | NA | 6.94E-11 | 9.17E-12 | 1.36E-02 | 7.04E-11 | 7.01E-12 | 9.96E-03 |
| SFRS12IP1 | SREK1 interacting protein 1 | 1.00E-10 | 1.00E-09 | 1 | 1.40E-10 | 1.97E-11 | 1.41E-02 | 1.44E-10 | 1.34E-11 | 9.32E-03 |
| APBB1IP | amyloid beta precursor protein binding family B member 1 interacting protein | 3.00E-10 | 1.71E-09 | 0.571488693 | 2.84E-10 | 2.90E-11 | 1.02E-02 | 2.98E-10 | 1.64E-11 | 5.49E-03 |
| SERPINB6 | serpin family B member 6 | 3.00E-10 | 1.71E-09 | 0.571488693 | 5.16E-10 | 5.01E-11 | 9.71E-03 | 5.35E-10 | 4.70E-10 | 8.63E-02 |
| PALM | paralemmin | 5.00E-10 | 2.61E-09 | 0.522232968 | 8.32E-10 | 1.08E-10 | 1.30E-02 | 8.15E-10 | 7.22E-11 | 8.86E-03 |
| ABCB1 | ATP binding cassette subfamily B member 1 | 8.00E-10 | 2.73E-09 | 0.340824905 | 8.85E-10 | 1.22E-10 | 1.39E-02 | 8.97E-10 | 8.74E-11 | 9.64E-03 |
| CCNB2 | cyclin B2 | 1.60E-09 | 4.20E-09 | 0.262322571 | 1.21E-09 | 1.39E-10 | 1.15E-02 | 1.26E-09 | 7.48E-11 | 5.94E-03 |
| ARC | activity regulated cytoskeleton associated protein | 1.80E-09 | 4.11E-09 | 0.228584567 | 1.57E-09 | 9.89E-11 | 6.30E-03 | 1.54E-09 | 5.99E-11 | 3.89E-03 |
| SNX19 | sorting nexin 19 | 1.81E-08 | 1.40E-08 | 0.077616668 | 1.99E-08 | 6.37E-10 | 3.20E-03 | 1.90E-08 | 3.31E-10 | 1.74E-03 |
| MIA3 | MIA SH3 domain ER export factor 3 | 4.36E-08 | 2.26E-08 | 0.051905936 | 4.38E-08 | 2.93E-09 | 6.70E-03 | 4.28E-08 | 1.18E-09 | 2.76E-03 |
| ARL15 | ADP ribosylation factor like GTPase 15 | 8.55E-08 | 3.27E-08 | 0.038212233 | 8.61E-08 | 7.56E-09 | 8.30E-03 | 8.64E-08 | 4.55E-09 | 5.27E-03 |
| PASK | PAS domain containing serine/threonine kinase | 1.81E-07 | 4.32E-08 | 0.023830639 | 1.85E-07 | 1.18E-08 | 6.40E-03 | 1.81E-07 | 5.77E-09 | 3.19E-03 |
| SCN11A | sodium voltage-gated channel alpha subunit 11 | 6.28E-07 | 7.62E-08 | 0.012148046 | 6.15E-07 | 6.00E-08 | 9.30E-03 | 6.09E-07 | 3.20E-08 | 5.26E-03 |
| PLAU | plasminogen activator, urokinase | 6.65E-07 | 8.16E-08 | 0.012273424 | 6.68E-07 | 2.74E-08 | 4.10E-03 | 6.45E-07 | 1.88E-08 | 2.91E-03 |
| FABP5L3 | fatty acid binding protein 5 pseudogene 3 | 6.94E-07 | 8.06E-08 | 0.011615522 | 6.84E-07 | 1.71E-08 | 2.50E-03 | 6.93E-07 | 7.14E-09 | 1.03E-03 |
| PELI3 | pellino E3 ubiquitin protein ligase family member 3 | 2.79E-06 | 1.56E-07 | 0.005578934 | 2.74E-06 | 2.11E-07 | 7.70E-03 | 2.85E-06 | 1.15E-07 | 4.04E-03 |
| AQPEP | laeverin | 3.98E-06 | 1.75E-07 | 0.004404336 | 4.06E-06 | 2.07E-07 | 5.10E-03 | 3.93E-06 | 1.27E-07 | 3.23E-03 |
| APOM | apolipoprotein M | 5.93E-06 | 2.39E-07 | 0.004021149 | 5.99E-06 | 1.32E-07 | 2.20E-03 | 5.89E-06 | 8.01E-08 | 1.36E-03 |

**P (crude)**, **P (CE-Perm1)** and **P (CE-Perm1-S)**: the final estimated p-value from each procedure, which is the average of the 100 runs of each procedure.

SD (crude), SD (CE-Perm1) and SD (CE-Perm1-S): the sample standard deviation of the estimated p-values from the 100 runs of each procedure.

MCRE (crude), MCRE (CE-Perm1) and MCRE (CE-Perm1-S): the Monte Carlo relative error as defined in Section 3.1, which is calculated based on the 100 runs of each procedure.

**Supplementary Table S2.** Comparison of the *p*-values from the permutation tests and those from the Welch's t-test for the top 23 significant probe sets in the MRD microarray data

| Probe Set ID | Crude *P*-value (*MCRE*) | CE-Perm2 *P*-value (*MCRE*) | CE-Perm2-S *P*-value (*MCRE*) | Welch's *t*-test *P*-value | Rank by Welch's *t*-test *P*-value |
|---|---|---|---|---|---|
| 242747_at | $2.40\times10^{-9}$ ($2.72\times10^{-1}$) | $2.71\times10^{-9}$ ($2.91\times10^{-2}$) | $2.58\times10^{-9}$ ($1.37\times10^{-2}$) | $6.88\times10^{-10}$ | 1 |
| 1564310_a_at | $3.80\times10^{-9}$ ($2.08\times10^{-1}$) | $4.39\times10^{-9}$ ($1.94\times10^{-2}$) | $4.31\times10^{-9}$ ($1.02\times10^{-2}$) | $6.11\times10^{-4}$ | 448 |
| 201718_s_at | $3.20\times10^{-9}$ ($2.47\times10^{-1}$) | $4.41\times10^{-9}$ ($1.65\times10^{-2}$) | $4.17\times10^{-9}$ ($9.61\times10^{-3}$) | $6.48\times10^{-8}$ | 4 |
| 219032_x_at | $3.02\times10^{-8}$ ($8.54\times10^{-2}$) | $2.89\times10^{-8}$ ($3.19\times10^{-2}$) | $2.83\times10^{-8}$ ($1.47\times10^{-2}$) | $1.98\times10^{-4}$ | 241 |
| 201719_s_at | $6.82\times10^{-8}$ ($4.34\times10^{-2}$) | $7.16\times10^{-8}$ ($1.22\times10^{-2}$) | $7.19\times10^{-8}$ ($8.33\times10^{-3}$) | $7.78\times10^{-8}$ | 5 |
| 205429_s_at | $8.98\times10^{-8}$ ($4.52\times10^{-2}$) | $8.67\times10^{-8}$ ($5.78\times10^{-3}$) | $8.74\times10^{-8}$ ($4.76\times10^{-3}$) | $6.49\times10^{-5}$ | 133 |
| 1553380_at | $1.12\times10^{-7}$ ($4.62\times10^{-2}$) | $1.07\times10^{-7}$ ($9.81\times10^{-3}$) | $1.03\times10^{-7}$ ($6.85\times10^{-3}$) | $3.61\times10^{-4}$ | 330 |
| 207426_s_at | $1.65\times10^{-7}$ ($3.67\times10^{-2}$) | $1.58\times10^{-7}$ ($1.65\times10^{-2}$) | $1.52\times10^{-7}$ ($8.76\times10^{-3}$) | $2.42\times10^{-5}$ | 74 |
| 209286_at | $1.76\times10^{-7}$ ($3.49\times10^{-2}$) | $1.73\times10^{-7}$ ($1.46\times10^{-2}$) | $1.78\times10^{-7}$ ($7.34\times10^{-3}$) | $1.06\times10^{-3}$ | 603 |
| 221841_s_at | $2.14\times10^{-7}$ ($2.84\times10^{-2}$) | $2.00\times10^{-7}$ ($9.05\times10^{-3}$) | $2.08\times10^{-7}$ ($6.57\times10^{-3}$) | $1.12\times10^{-5}$ | 51 |
| 227336_at | $4.17\times10^{-7}$ ($2.12\times10^{-2}$) | $4.27\times10^{-7}$ ($5.81\times10^{-3}$) | $4.22\times10^{-7}$ ($5.02\times10^{-3}$) | $2.02\times10^{-6}$ | 17 |
| 225685_at | $4.75\times10^{-7}$ ($2.07\times10^{-2}$) | $4.89\times10^{-7}$ ($6.18\times10^{-3}$) | $4.78\times10^{-7}$ ($4.39\times10^{-3}$) | $2.25\times10^{-3}$ | 899 |
| 213358_at | $6.30\times10^{-7}$ ($1.75\times10^{-2}$) | $6.16\times10^{-7}$ ($7.66\times10^{-3}$) | $6.14\times10^{-7}$ ($4.27\times10^{-3}$) | $5.51\times10^{-5}$ | 120 |
| 219990_at | $6.57\times10^{-7}$ ($1.60\times10^{-2}$) | $6.60\times10^{-7}$ ($9.52\times10^{-3}$) | $6.63\times10^{-7}$ ($3.59\times10^{-3}$) | $5.87\times10^{-6}$ | 31 |
| 204562_at | $6.78\times10^{-7}$ ($1.76\times10^{-2}$) | $6.70\times10^{-7}$ ($5.97\times10^{-3}$) | $6.65\times10^{-7}$ ($4.38\times10^{-3}$) | $3.81\times10^{-6}$ | 24 |
| 213817_at | $8.91\times10^{-7}$ ($1.48\times10^{-2}$) | $8.71\times10^{-7}$ ($5.63\times10^{-3}$) | $8.79\times10^{-7}$ ($3.04\times10^{-3}$) | $7.09\times10^{-4}$ | 481 |
| 201710_at | $8.89\times10^{-7}$ ($1.44\times10^{-2}$) | $8.95\times10^{-7}$ ($5.53\times10^{-3}$) | $8.93\times10^{-7}$ ($2.26\times10^{-3}$) | $6.74\times10^{-6}$ | 33 |
| 232539_at | $9.79\times10^{-7}$ ($1.38\times10^{-2}$) | $9.58\times10^{-7}$ ($6.01\times10^{-3}$) | $9.62\times10^{-7}$ ($2.89\times10^{-3}$) | $7.11\times10^{-6}$ | 36 |
| 218589_at | $1.36\times10^{-6}$ ($1.23\times10^{-2}$) | $1.37\times10^{-6}$ ($5.15\times10^{-3}$) | $1.35\times10^{-6}$ ($2.63\times10^{-3}$) | $2.94\times10^{-5}$ | 83 |
| 218899_s_at | $1.54\times10^{-6}$ ($1.28\times10^{-2}$) | $1.57\times10^{-6}$ ($4.29\times10^{-3}$) | $1.55\times10^{-6}$ ($2.24\times10^{-3}$) | $9.66\times10^{-5}$ | 159 |
| 225688_s_at | $2.04\times10^{-6}$ ($9.12\times10^{-3}$) | $2.06\times10^{-6}$ ($6.36\times10^{-3}$) | $2.05\times10^{-6}$ ($2.19\times10^{-3}$) | $4.54\times10^{-5}$ | 107 |
| 242051_at | $5.66\times10^{-6}$ ($5.58\times10^{-3}$) | $5.66\times10^{-6}$ ($5.02\times10^{-3}$) | $5.67\times10^{-6}$ ($2.87\times10^{-3}$) | $6.62\times10^{-5}$ | 134 |
| 220448_at | $7.03\times10^{-6}$ ($5.31\times10^{-3}$) | $7.08\times10^{-6}$ ($4.79\times10^{-3}$)* | $7.07\times10^{-6}$ ($2.53\times10^{-3}$) | $7.08\times10^{-4}$ | 480 |

*For this probe set, there is one single run of CE-Perm2 where the adaptive updating step does not converge. The *p*-value and *MCRE* presented are based on the converged 99 runs.